\documentclass{JHEP3}
\usepackage{amsmath,multirow}
\newcommand{\ten}{{\hskip1.0mm\cdot\hskip1.0mm}}
\newcommand{\dt}{{\circle*{.3}}}
\newcommand{\vl}{{\linethickness{1.6pt}\line(0,1){1}}}
\newcommand{\hl}{{\linethickness{1.6pt}\line(1,0){1}}}
\newcommand{\dl}{{\linethickness{0.2pt}\line(1,1){1}}}
\newcommand{\num}[1]{{{\makebox(-.4,.7){\tiny$#1$}}}}

\title{Self-dual Strings and 2D SYM}

\author{Kazuo Hosomichi$^a$ and Sungjay Lee$^b$\\
$^a$Yukawa Institute for Theoretical Physics,
    Kyoto University, Japan \\
$^b$Enrico Fermi Institute and Department of Physics,
    University of Chicago, USA
\vskip2mm
E-mail:
\email{hosomiti@yukawa.kyoto-u.ac.jp},
\email{sjlee79@uchicago.edu}
}

\abstract{
We study the system of M2-branes suspended between parallel M5-branes
using ABJM model with a natural half-BPS boundary condition. For small
separation between M5-branes, the worldvolume theory is shown to reduce
to a 2D ${\cal N}=(4,4)$ super Yang-Mills theory with some similarity to
$q$-deformed Yang-Mills theory. The gauge coupling is related to the
position of the branes in an interesting manner. The theory is
considerably different from the 2D theory proposed for multiple
``M-strings''. We make a detailed comparison of elliptic genus of
the two descriptions and find only a partial agreement.
}

\preprint{EFI-14-11\\YITP-14-43}


\begin{document}

\section{Introduction and Summary}

The (2,0) theory in six dimensions, which describes the many-body
dynamics of M5-branes, is the unique superconformal field theory with
maximal supersymmetry and in the highest possible dimensions. Despite its
unique character, our understanding of this theory is far from satisfactory
mainly due to the lack of Lagrangian description, although many
interesting discovery have been made with regard to its property
under compactifications.

As one of its distinct features, the (2,0) theory has tensionless
self-dual strings as its fundamental degree of freedom. On the Coulomb
branch of the theory where the M5-branes are placed at distances from
one another, these strings are described by open M2-branes suspended
between the M5-branes \cite{Haghighat:2013gba}. The BPS spectrum of the
worldsheet theory of the self-dual strings have been studied in
\cite{Haghighat:2013gba,Haghighat:2013tka} in a dual IIA Calabi-Yau
background using the refined topological vertex formalism. In this paper
we study the same system using the ABJM model \cite{Aharony:2008ug}.

In Section \ref{sec:setup} we present the brane configuration in M-theory and
summarize the symmetry of the system. In Section \ref{sec:ABJM} we
review and extend the analysis of half-BPS boundary conditions of the
ABJM model \cite{Berman:2009xd}, focusing particularly on the one provided by
M5-branes. We also study the M-theoretic analogue of the Nahm equation
for ground states of multiple M2-branes suspended between M5-branes.
It has been known that the M2-branes develop Nahm-like pole near their
endpoints at M5-branes, but we find this occurs only when the system of
M2 and M5-branes is at the origin of the transverse space
$\mathbb C^2/\mathbb Z_k$.

In Section \ref{sec:worldsheet} we take the separation between M5-branes
to be very small and reduce the worldvolume theory of suspended
M2-branes to two dimensions. The resulting theory, which we call {\it the
ABJM slab}, is generically the ${\cal N}=(4,4)$ super Yang-Mills
theory with R-symmetry $SU(2)^3$, with the gauge coupling $g^2$
determined from the distance between the brane composite and the $\mathbb Z_k$
orbifold singularity. Since it is a dimensional reduction of a
Chern-Simons gauge theory, the kinetic term for the gauge field takes
the first order form $\text{Tr}(\phi F_{01}+g^2\phi^2)$, and the adjoint
scalar $\phi$ is subject to a certain periodicity. The gauge coupling
vanishes when the branes are right at the orbifold singularity, and then
the theory simplifies to a topological Yang-Mills theory coupled to
adjoint matters with ${\cal N}=(4,4)$ supersymmetry and an enhanced
R-symmetry $SU(2)^4$. There is some similarity between the
theory we obtained and the so-called $q$-deformed Yang-Mills theory
\cite{Aganagic:2004js}; some detailed comparison is made in Section
\ref{sec:qYM}. Also, our theory is rather different from the theory
proposed in \cite{Haghighat:2013tka} based on brane construction in a
dual type IIA framework. This model will be called {\it the IIA brane
model} in this paper and is reviewed in Sections \ref{sec:IIAbrane} and
\ref{sec:IIAquiver}.

In Section \ref{sec:elliptic} we study the elliptic genus, namely the
$T^2$ partition function, for the two descriptions of multiple self-dual
strings as a function of fugacity variables $\epsilon_1,\epsilon_2,m$
parametrizing twists in the periodicity of fields. It was claimed in
\cite{Haghighat:2013tka} that the formula for the elliptic genus
obtained from refined topological vertex \cite{Haghighat:2013gba} can be
reproduced using the IIA brane model. In Section \ref{sec:EG-IIA} we
quickly confirm this using the idea of Higgs branch localization. Then
the elliptic genus of the ABJM slab is studied in Section
\ref{sec:EGABJM}. There it is argued that the cases $g^2>0$ and $g^2=0$
require different treatment, and the former corresponds to the special
choice $m=\pm\frac12(\epsilon_1+\epsilon_2)$ of the fugacity parameters.
Moreover it is shown that the elliptic genera of the IIA brane model and
the ABJM slab behave differently in the limit
$m\to\pm\frac12(\epsilon_1+\epsilon_2)$. On the other hand we show,
using the Jeffrey-Kirwan residue formula
\cite{Benini:2013nda,Benini:2013xpa}, that the elliptic genus for
$g^2>0$ is that of ${\cal N}=(4,4)$ super Yang-Mills theory and is
expressed as a Young diagram sum.

We conclude with a few remarks in Section \ref{sec:conclusion}. The
Appendix \ref{sec:appendix} shows an off-shell supersymmetric
formulation of ABJM model on a slab for the simplest case $N=1$.

\section{The setup}\label{sec:setup}

In this paper we study the system of parallel M2-branes suspended
between parallel M5-branes using the ABJM model \cite{Aharony:2008ug}.

The coordinates $(x_0,x_1,x_2)$ are used for the M2-brane worldvolume
direction, and the transverse direction is parametrized either by
eight real coordinates $(x_3,\cdots,x_9,x_\natural)$ or four complex
coordinates $(z_1,\cdots,z_4)\equiv(x_3+ix_4,\cdots,x_9+ix_\natural)$.
The ABJM model describes the transverse geometry $\mathbb C^4/\mathbb Z_k$,
where the $\mathbb Z_k$ orbifold rotates the phase of $z_i$
simultaneously. In addition, we introduce M5-branes with the worldvolume
direction $(013456)$, mutually separated in the $x_2$ direction. We
mostly focus on the case with two M5-branes, one at $x_2=0$ and the
other at $x_2=L$, with $N$ M2-branes suspended between them. They are
also labeled by the common position in the remaining 4 transverse
direction $(789\natural)$. Our brane configuration is summarized in the
table below.
\begin{center}
 \begin{tabular}{|c|ccccccccccc|}
 \hline
   & 0 & 1 & 2 & 3 & 4 & 5 & 6 & 7 & 8 & 9 & $\natural$ \\
 \hline
  M2  &$-$&$-$&$\dashv$&  &  &  &  &  &  &  &  \\
 \hline
  M5  &$-$&$-$& &$-$&$-$&$-$&$-$&  &  &  &  \\
 \hline
 \end{tabular}
\end{center}
Note that, if the branes are at $x_7=\cdots=x_\natural=0$, each M5-brane
has the worldvolume $\mathbb R^{1,1}(012)\times \mathbb R^4/\mathbb Z_k(3456)$.
Otherwise their worldvolume is $\mathbb R^{1,1}\times \mathbb R^4$.

Let us study how many supersymmetry is preserved by this system, by
using the set of 11D Gamma matrices $\Gamma^a$. In our convention
$\Gamma^0$ is anti-Hermitian and the rest are all Hermitian, and they
satisfy
\begin{equation}
 \Gamma^0\Gamma^1\cdots\Gamma^9\Gamma^\natural={\bf 1}_{32\times32}.
\end{equation}
We denote by $Q$ the 32-component supersymmetry on 11D flat spacetime.

First, the M2-branes (012) are half-BPS objects preserving the
supersymmetry characterized by
\begin{equation}
 Q=\Gamma^{012}Q=\Gamma^{3456789\natural}Q.
\end{equation}
The $\mathbb Z_k$ orbifolding for $k\ge3$ reduces further the unbroken
supersymmetry to those satisfying
\begin{equation}
 \exp\Big\{
 \frac{\pi}{k}(\Gamma^{34}+\Gamma^{56}+\Gamma^{78}+\Gamma^{9\natural})
 \Big\}Q=Q.
\end{equation}
Working with the basis of 11D spinors diagonalizing
$(i\Gamma^{34},i\Gamma^{56},i\Gamma^{78},i\Gamma^{9\natural})$, this
projects out the 2/8 of the unbroken supersymmetry characterized by the
eigenvalues $({+}{+}{+}{+})$ and $({-}{-}{-}{-})$. One is thus left with
12 supercharges or 3D ${\cal N}=6$ supersymmetry on the M2-brane
worldvolume in the absence of the M5-branes.

The M5-branes (013456) further reduce the unbroken supersymmetry to
those satisfying
\begin{equation}
 Q=\Gamma^{013456}Q.
\end{equation}
They also introduce the boundary to the M2-brane worldvolume. The
remaining supercharges are the following six, characterized by the
eigenvalues
\begin{eqnarray}
\lefteqn{
(\Gamma^{01},\Gamma^2,i\Gamma^{34},i\Gamma^{56},
i\Gamma^{78},i\Gamma^{9\natural})
} \nonumber \\
 &=&
({+}{+}{+}{-}{+}{-}),({+}{+}{+}{-}{-}{+}),
({+}{+}{-}{+}{+}{-}),({+}{+}{-}{+}{-}{+}),\nonumber \\ &&
({-}{-}{+}{+}{-}{-}),({-}{-}{-}{-}{+}{+}).
\end{eqnarray}
The M2-branes suspended between the M5-branes have the worldvolume
$\mathbb R^{1,1}\times(\text{interval})$, and are described at low
energy by a 2D theory. If one identifies $\Gamma^{01}$ with the
chirality of 2D spinors, the resulting theory is expected to have 4
chiral and 2 anti-chiral 2D supercharges. The M5-branes also break the
$SU(4)$ R-symmetry of the 3D ${\cal N}=6$ supersymmetric theory to
$SU(2)\times SU(2)\times U(1)$. The latter is in agreement with the
R-symmetry of the 2D ${\cal N}=(4,2)$  supersymmetric systems.

We note here that one can introduce M5-branes $(01789\natural)$ or
9-branes $(013456789\natural)$ of appropriate orientation without
breaking supersymmetry further.

\section{ABJM model on a slab}\label{sec:ABJM}

The system of parallel M2-branes probing the transverse geometry
$\mathbb C^4/\mathbb Z_k$ is described by the ABJM model
\cite{Aharony:2008ug}. To describe M2-branes suspended between
M5-branes, we consider the model on a slab
$\mathbb R^{1,1}\times\text{(interval)}$ with suitable boundary
conditions. We summarize here the construction of the model.

We work in the 3D Minkowski spacetime with $({-}{+}{+})$ signature. Spinors
are 2-component quantities on which $2\times 2$ gamma matrices act as
\begin{equation}
 \gamma_m\psi \equiv (\gamma_m)_\alpha^{~\beta}\psi_\beta.\quad
 (\alpha,\beta=\pm)
\end{equation}
Bilinear products of spinors are defined using the real anti-symmetric
matrix $C^{\alpha\beta}$ as follows.
\begin{equation}
 \xi\psi \equiv C^{\alpha\beta}\xi_\alpha\psi_\beta,\quad
 \xi\gamma_m\psi \equiv (C\gamma_m)^{\alpha\beta}\xi_\alpha\psi_\beta.
\end{equation}
We use
\begin{equation}
\def\arraystretch{.75}
 C^{\alpha\beta}=\Big(\begin{array}{cc} 0&1 \\-1&0 \end{array}\Big),\quad
 \gamma^0=\Big(\begin{array}{cc} 0&1 \\-1&0 \end{array}\Big),\quad
 \gamma^1=\Big(\begin{array}{cc} 0&1 \\ 1&0 \end{array}\Big),\quad
 \gamma^2=\Big(\begin{array}{cc} 1&0 \\ 0&-1\end{array}\Big),
\def\arraystretch{1.0}
\end{equation}
so that $(C\gamma^m)^{\alpha\beta}$ are all real symmetric and moreover
$\gamma^0\gamma^1\gamma^2=1$. Also, the components $\psi_+,\psi_-$ of a
spinor $\psi$ are for the eigenvalues of the chirality
$\gamma^2=\gamma^{01}=\pm1$ upon dimensional reduction to 2D.

\subsection{The bulk theory}

The ABJM model is a 3D ${\cal N}=6$ supersymmetric
$U(N)_k\times U(N)_{-k}$ Chern-Simons theory coupled to bifundamental
scalars $Z_a$ and spinors $\Psi^a$, where $a=1,\cdots,4$ is the $SU(4)$
R-symmetry index. All the fields are $N\times N$ matrix-valued, and the
Lagrangian takes the form
\begin{equation}
 {\cal L} ~=~
 {\cal L}_\text{CS}+{\cal L}_\text{kin}
+{\cal L}_\text{yuk}+{\cal L}_\text{pot},
\label{LABJM}
\end{equation}
with
\begin{eqnarray}
 {\cal L}_\text{CS} &=&
 \frac k{4\pi}\epsilon^{mnp}\,\text{Tr}\Big(
   A_m\partial_nA_p-\frac{2i}3A_mA_nA_p
  -\tilde A_m\partial_n\tilde A_p+\frac{2i}3\tilde A_m\tilde A_n\tilde A_p
 \Big),
 \nonumber \\
 {\cal L}_\text{kin} &=&
 \text{Tr}\Big(
  i\bar\Psi_a\gamma^mD_m\Psi^a
 -D_m\bar Z^aD^mZ_a
 \Big),
 \nonumber \\
 {\cal L}_\text{yuk} &=&
 \frac{2\pi i}k\text{Tr}\Big(
  \Psi^a\bar\Psi_aZ_b\bar Z^b-\bar\Psi_a\Psi^a\bar Z^bZ_b
 -2Z_a\bar Z^b\Psi^a\bar\Psi_b+2\bar Z^aZ_b\bar\Psi_a\Psi^b
 \nonumber \\ && \hskip10mm
 +\epsilon^{abcd}Z_a\bar\Psi_bZ_c\bar\Psi_d
 -\epsilon_{abcd}\bar Z^a\Psi^b\bar Z^c\Psi^d
 \Big),
 \nonumber \\
 {\cal L}_\text{pot} &=&
 \frac{4\pi^2}{3k^2}\text{Tr}\Big(
  4Z_a\bar Z^bZ_c\bar Z^aZ_b\bar Z^c
 -6Z_a\bar Z^bZ_b\bar Z^aZ_c\bar Z^c
 \nonumber \\ && \hskip11mm
  +Z_a\bar Z^aZ_b\bar Z^bZ_c\bar Z^c
  +Z_a\bar Z^bZ_b\bar Z^cZ_c\bar Z^a
 \Big)\,.
\end{eqnarray}
In our convention the covariant derivative acting on bifundamental
fields is defined by, for example,
$DZ_a\equiv{\rm d}Z_a-iAZ_a+iZ_a\tilde A$. Therefore the gauge field
strength is defined by $F={\rm d}A-iA^2$ and similarly for $\tilde F$.

The ${\cal N}=6$ supersymmetry is parametrized by a spinor $\xi_{ab}$
satisfying $\xi_{ab}=-\xi_{ba}$ and the reality condition
$(\xi_{ab})^\dagger \equiv \xi^{ab} = \frac12\epsilon^{abcd}\xi_{cd}$.
The transformation rule reads
\begin{eqnarray}
\delta Z_a &=& i\xi_{ab}\Psi^b,\nonumber \\
\delta\bar Z^a &=& i\xi^{ab}\bar\Psi_b, \nonumber \\
\delta\Psi^a &=& \gamma^m\xi^{ab}D_mZ_b + \xi^{bc}W_{bc}^a,\nonumber \\
\delta\bar\Psi_a &=&
 \gamma^m\xi_{ab}D_m\bar Z^b+\xi_{bc}\bar W^{bc}_a,\nonumber \\
\delta A_m &=& -\frac{2\pi}k(\xi_{ab}\gamma_m\Psi^a\bar Z^b
  +Z_a\xi^{ab}\gamma_m\bar\Psi_b), \nonumber \\
\delta\tilde A_m &=& +\frac{2\pi}k(\xi^{ab}\gamma_m\bar\Psi_aZ_b
  +\bar Z^a\xi_{ab}\gamma_m\Psi^b),
\label{SUSYABJM}
\end{eqnarray}
where
\begin{eqnarray}
 W^a_{bc} &=& +\frac\pi k
 \Big\{2Z_b\bar Z^aZ_c
 +\delta^a_b(Z_c\bar Z^dZ_d-Z_d\bar Z^dZ_c)
 \Big\}-(b\leftrightarrow c),
 \nonumber \\
 \bar W_a^{bc} &=& -\frac\pi k
 \Big\{2\bar Z^bZ_a\bar Z^c
 +\delta_a^b(\bar Z^cZ_d\bar Z^d-\bar Z^dZ_d\bar Z^c)
 \Big\}-(b\leftrightarrow c).
\end{eqnarray}
Note that the scalar potential can be expressed as
$-{\cal L}_\text{pot}=\frac23\text{Tr}(W^a_{bc}\bar W^{bc}_a)$, so it is
manifestly positive definite.

\subsection{Boundary condition at M5-branes}

To describe the M2-branes suspended between two M5-branes (013456) at
$x_2=0$ and $x_2=L$, we need to consider the ABJM model with suitable
boundary conditions on fields. The correct boundary condition should
preserve 6 of the 12 supercharges $Q_{ab\alpha}$ and the
$SU(2)\times SU(2)\times U(1)$ subgroup of the $SU(4)$ R-symmetry group.
The four complex scalars $Z_a$ are divided into two pairs by the
M5-brane boundary condition. Namely,
\begin{equation}
\begin{array}{cccl}
 Z_I & (I=1,2) &:& \text{longitudinal to the M5-branes,} \\
 Z_A & (A=3,4) &:& \text{transverse to the M5-branes.}
\end{array}
\end{equation}
These two pairs are
assigned opposite $U(1)$ R-charges, and are transformed by the two
respective $SU(2)$ factors. In short, ${\bf 4}$ of $SU(4)$ decomposes
into $\bf (2,1)_1\oplus(1,2)_{-1}$ of the reduced R-symmetry
$SU(2)\times SU(2)\times U(1)$.

From the discussion in the previous section, the unbroken supercharges
should form a 2D ${\cal N}=(4,2)$ superalgebra. Therefore the unbroken
supercharges are
\begin{equation}
 Q_{13+},~Q_{14+},~Q_{23+},~Q_{24+},~Q_{12-},~Q_{34-}\,.
\label{susypres}
\end{equation}
The four chiral supercharges furnish a $\bf 4$ of $SO(4)$, while the two
anti-chiral supercharges are $SO(4)$ singlet but carry nonzero $U(1)$ charges.
In order for the supercharge $Q_{ab\alpha}$ to be preserved, the
normal component of the corresponding supercurrent has to vanish, namely
${\cal J}^2_{ab\alpha}=0$. Therefore we put
\begin{equation}
 \gamma^2{\cal J}^2_{ab} ~=~ P_a^{~c}P_b^{~d}{\cal J}^2_{cd},\quad
 P=\text{diag}(+1,+1,-1,-1).
\end{equation}
Using the doublet indices $I,J,\cdots$ and $A,B,\cdots$ for the two
$SU(2)$ R-symmetry, this can be rewritten as follows,
\begin{equation}
 {\cal J}^2_{IA+}=
 {\cal J}^2_{IJ-}=
 {\cal J}^2_{AB-}=0.
\label{BC-J}
\end{equation}
The supercurrent of the ABJM model can be obtained in the standard
manner. Using a spinor $\xi^{ab}=-\xi^{ba}$ satisfying also
$\xi^{ab}=\frac12\epsilon^{abcd}\xi_{cd}$, it can be expressed as
\begin{eqnarray}
 \xi^{ab}{\cal J}^m_{ab} &=&
 \xi^{ab}\text{Tr}
 (\gamma^n\gamma^m\bar\Psi_aD_nZ_b-\gamma^m\bar\Psi_cW^c_{ab})
 \nonumber \\ &&
+\xi_{ab}\text{Tr}
 (\gamma^n\gamma^m\Psi^aD_n\bar Z^b-\gamma^m\Psi^c\bar W_c^{ab}).
\end{eqnarray}

The boundary condition on supercurrents (\ref{BC-J}) must follow from
that on the gauge and matter fields. Let us first find out the boundary
condition on the fermions $\Psi^a,\bar\Psi_a$. It has to be given by an
$SU(2)\times SU(2)\times U(1)$ invariant linear equation in the fermions.
Moreover, the left-right asymmetry of the unbroken supercharges suggests
that the equation should involve $\gamma^2$. Thus the only possible form
one can think of is
\begin{equation}
 \gamma^2\Psi^a=\pm\Psi^bP_b^{~a},\quad
 \gamma^2\bar\Psi_a=\pm P_a^{~b}\bar\Psi_b.
\end{equation}
It turns out that the boundary condition corresponding to the M5-brane
(013456) corresponds to the choice of $-$ sign. For this choice, the
above boundary condition can be rewritten more explicitly as follows.
\begin{equation}
 \Psi^I_+=\Psi^A_-=0,\quad
 \bar\Psi_{I+}=\bar\Psi_{A-}=0.
\label{BC-PSI}
\end{equation}
The other choice of sign corresponds to the M5-brane
$(01789\natural)$. These boundary conditions, along with some other
half-BPS boundary conditions in the ABJM model, were studied in some
detail in \cite{Berman:2009xd}.

By combining (\ref{BC-J}) with (\ref{BC-PSI}), one obtains the boundary
condition on bosons. The normal components of supercurrent in
(\ref{BC-J}) are linear in the fermions, and we require the coefficients of
$\Psi^A_+,\Psi^I_-,\bar\Psi_{A+},\bar\Psi_{I-}$
to vanish on the boundary. This leads to the following conditions
\cite{Berman:2009xd}
\begin{eqnarray}
&&
 D_\mu Z_A=0,
\nonumber \\ &&
 D_y Z_I=\frac{2\pi}k
 (Z_I\bar Z^JZ_J-Z_J\bar Z^JZ_I),
\nonumber \\ &&
  Z_A\bar Z^IZ_B=Z_B\bar Z^IZ_A,\quad
  Z_I\bar Z^AZ_B=Z_B\bar Z^AZ_I,
\label{bcsc}
\end{eqnarray}
and their Hermite conjugates. Here the index $\mu$ takes $0$ and $1$,
and we denoted the coordinate $x_2$ by $y$ for convenience.
The first is the Dirichlet boundary condition for $Z_A=(Z_3,Z_4)$. They
are therefore constant along the boundary, and their boundary value
determines the position of the M5-brane (013456). The second is the
M-theoretic analogue of Nahm pole \cite{Gaiotto:2008sa}. The third
condition determines how the shape of Nahm pole is restricted by the
nonzero values of $Z_A$. For example, assume $Z_A$ takes the form
\begin{equation}
 Z_A=\left(\begin{array}{ccc}
 u^{(1)}_A\cdot{\bf 1}_{n_1\times n_1} && \\
 & u^{(2)}_A\cdot{\bf 1}_{n_2\times n_2} & \\
 && \ddots \end{array}
\right)\,.
\label{ZAbv}
\end{equation}
Then $Z_I=(Z_1,Z_2)$ can take nonzero values only within the diagonal
blocks of size $n_1\times n_1, n_2\times n_2$ and so on. This
corresponds to the situation with several M5-branes at different points
in $\mathbb C^2/\mathbb Z_k$. The $i$-th M5-brane is at $z_A=u_A^{(i)}$
and there are $n_i$ M2-branes ending on it.

The boundary conditions on gauge fields follow from the Dirichlet
boundary condition on $Z_A$. By differentiating the first equation in
(\ref{bcsc}) one more time one finds
\begin{equation}
 F_{\mu\nu}Z_A=Z_A\tilde F_{\mu\nu},
\label{bc-F}
\end{equation}
which glues the two $U(N)$ gauge fields. For example, if $Z_A$
takes the block diagonal form (\ref{ZAbv}) along the boundary, the gauge
fields can take nonzero values only within the diagonal blocks, and
moreover $A_\mu=\tilde A_\mu$ for each block. The gauge symmetry
$U(N)\times U(N)$ is thus broken to
$U(n_1)_\text{diag}\times U(n_2)_\text{diag}\times\cdots$ on the boundary.
Note that this boundary condition on gauge fields also resolves the
problem of gauge non-invariance of Chern-Simons theory with boundary. Namely
the ABJM action is gauge invariant only up to the surface term,
\begin{equation}
 \delta S~=~ \frac k{4\pi}\int_{\text{boundary}}\text{Tr}
 (\alpha {\rm d}A-\tilde\alpha {\rm d}\tilde A),
\label{CS-noinv}
\end{equation}
but it vanishes if $A=\tilde A,~\alpha=\tilde\alpha$ are imposed on the
boundary.

More general solution to (\ref{bc-F}) would be that the gauge fields
$A_\mu$ and $\tilde A_\mu$ are unitary equivalent. Namely, one could generalize
(\ref{ZAbv}) so that the $i$-th diagonal block of $Z_A$ is $u_A^{(i)}$
times a unitary matrix $U^{(i)}\in SU(n_i)$, and relate the
corresponding blocks of the gauge fields by
$A_\mu^{(i)}U^{(i)}=U^{(i)}\tilde A_{\mu}^{(i)}$. One can regard the
$U^{(i)}$'s as gauge redundancy and choose them to be all identity.
The $U(1)$ part of the gauge group $U(n_i)$ rotates the phase of
$u_A^{(i)}$, but this $U(1)$ rotation for general angle is not a
symmetry of the ABJM model due to monopoles. As a consequence, the
parameter $u_A^{(i)}$ is subject to the identification only under the
$\mathbb Z_k$ rotation, not the full $U(1)$ rotation.

The supersymmetry preserved by the boundary condition is
expressed as $\xi^{ab}Q_{ab}$ with
\begin{equation}
 \gamma^2\xi_{ab}= P_a^{~c}P_b^{~d}\xi_{cd}\,.
\label{halfBPS}
\end{equation}
More explicitly, the unbroken supersymmetry is parametrized by
$\xi^{IJ}_+,\xi^{AB}_+$ and $\xi^{IA}_-$.
It is straightforward to check that the boundary conditions on bosons
and fermions transform among themselves under the unbroken
supersymmetry. The only additional condition is that, if $Z_A$ takes the
block-diagonal form (\ref{ZAbv}), then the boundary value of all the
fields are restricted to take the same block-diagonal form.

\vskip2mm

Before closing this subsection, let us comment on the important special
case $Z_A=0$. The condition (\ref{bc-F}) is then trivially satisfied,
and one does not need to identify the two $U(N)$ gauge fields along the
boundary. However, if one is to relax the condition $A_\mu=\tilde A_\mu$
along the boundary, then the gauge invariance has to be ensured by
some other means, for example by introducing boundary chiral fermions
(or WZW models as was proposed in \cite{Chu:2009ms}). Namely, by
coupling the boundary fermions of one chirality to $A_\mu$ and the other
to $\tilde A_\mu$, one gets a desired gauge anomaly which cancels
(\ref{CS-noinv}). It is not clear to us whether the 2D ${\cal N}=(4,2)$
supersymmetry allows multiplets containing such chiral fermions. In this
paper we stick to the gluing condition $A_\mu=\tilde A_\mu$ even when
$Z_A=0$, and leave the study of possible boundary degree of freedom as a
future problem.

\subsection{Boundary term}

We also need to check that the variational problem is well defined for
the ABJM model with boundary, namely solving the bulk Euler-Lagrange
equation and the boundary condition should always lead to stationary phase
configurations. By a careful study of this requirement one finds that
the bulk ABJM action has to be supplemented with a suitable boundary term.

The variation of the action of the ABJM model on a slab
$0\le y\le L$ with Lagrangian (\ref{LABJM}) takes the form
\begin{equation}
 \delta S ~=~ \int {\rm d}^3x \big(\text{Euler-Lagrange eqn}\big)
 -\int_{y=L}\hskip-0mm {\rm d}^2x\, I_\text{bd}
 +\int_{y=0}\hskip-0mm {\rm d}^2x\, I_\text{bd},
\end{equation}
where the surface term $I_\text{bd}$ reads
\begin{equation}
 I_\text{bd} =
 \text{Tr}\Big(-\frac k{4\pi}\varepsilon^{\mu\nu}
 (\delta A_\mu A_\nu-\delta\tilde A_\mu\tilde A_\nu)
 -i\bar\Psi_a\gamma^2\delta\Psi^a
 +\delta\bar Z^aD_yZ_a+D_y\bar Z^a\delta Z_a \Big).
\end{equation}
Most of the terms in $I_\text{bd}$ vanish once the boundary conditions
on fields are taken into account, except the terms proportional to
$\delta Z_I,\delta\bar Z^I$. Using the Nahm pole boundary condition one finds
\begin{equation}
 I_\text{bd} =
 \frac{2\pi}k\text{Tr}
 \Big(\delta\bar Z^I(Z_I\bar Z^JZ_J-Z_J\bar Z^JZ_I)
 +\delta Z_I(\bar Z^JZ_J\bar Z^I-\bar Z^IZ_J\bar Z^J) \Big)
 = \delta {\cal L}_\text{bd}.
\end{equation}
Namely the surface term is nonvanishing but can be canceled by the
boundary Lagrangian
\begin{equation}
 {\cal L}_\text{bd} = \frac\pi k\text{Tr}
 (\bar Z^IZ_I\bar Z^JZ_J-Z_I\bar Z^IZ_J\bar Z^J).
\end{equation}
Thus the total action must include the boundary contribution
\begin{equation}
 S_\text{tot} = S_\text{bulk}
 +\int_{y=L}{\rm d}^2x{\cal L}_\text{bd}
 -\int_{y=0}{\rm d}^2x{\cal L}_\text{bd}.
\label{Lbd}
\end{equation}

\subsection{Nahm equation}

The word Nahm pole originally refers to a characteristic singular
behavior of the solutions of Nahm equation at the boundary. As is well
known, Nahm equation is a system of ordinary differential equations for
a triplet of matrix valued functions \cite{Nahm:1979yw}. The solutions
are called Nahm data and are used to construct monopole configurations in
Yang-Mills theory. In superstring theory, Nahm equation arises as the
BPS equation on the worldvolume of D2-branes ending on D4-branes
\cite{Diaconescu:1996rk}. An analogue of Nahm equation is known for
M2-branes ending on M5-branes; see \cite{Basu:2004ed,Nogradi:2005yk} for
earlier work and
\cite{Terashima:2008sy,Gomis:2008vc,Hanaki:2008cu,Nosaka:2012tq,Sakai:2013gga}
for the work based on the ABJM model. Let us revisit this equation here.

We solve the BPS equation
$\delta\Psi^a=\delta\bar\Psi_a=0$ with the constraint (\ref{halfBPS}) on
supersymmetry parameter $\xi_{ab}$. The solution should also preserve the
$SO(1,1)\times SU(2)\times SU(2)\times U(1)$ subgroup of the 3D Lorentz
and R-symmetries. The condition on bosonic fields is given by
$D_\mu Z_I=D_\mu Z_A=0$, namely the fields depend only on
$y\equiv x_2$, and
\begin{eqnarray}
 &&
 D_yZ_I ~=~ \frac{2\pi}k(Z_I\bar Z^JZ_J-Z_J\bar Z^JZ_I),
 \nonumber \\
 &&
 D_yZ_A ~=~ \frac{2\pi}k(Z_A\bar Z^BZ_B-Z_B\bar Z^BZ_A),
 \nonumber \\
 &&
 Z_I\bar Z^AZ_J=Z_J\bar Z^AZ_I,
 \quad
 Z_I\bar Z^JZ_A=Z_A\bar Z^JZ_I,
 \nonumber \\
 &&
 Z_A\bar Z^IZ_B=Z_B\bar Z^IZ_A,\quad
 Z_A\bar Z^BZ_I=Z_I\bar Z^BZ_A.
\end{eqnarray}
The equations are symmetric under the exchange of the indices $I=1,2$
and $A=3,4$.

The first and the second equations are the Nahm equations for M2-branes
ending on M5-branes (013456) or $(01789\natural)$. It turns out,
however, that the solutions show nontrivial dependence on $y$ only in very
limited cases, since the additional algebraic equations in the
last two lines generically require that $Z_I,Z_A$ are all diagonal in a
suitable gauge. Let us quickly show this. We take a generic linear
combination of $Z_A$ which has no accidental zero eigenvalues or
degeneration of eigenvalues when it is gauge-rotated to a diagonal form.
Let us assume $Z_3$ is generic, and we gauge rotate it to the form
(\ref{ZAbv}) made of blocks which are proportional to identity matrices.
Then it follows from $Z_3\bar Z^3Z_I=Z_I\bar Z^3Z_3$ that $Z_I,\bar Z^I$
are also block diagonal, and moreover their $j$-th blocks have to commute with
each other as long as the $j$-th block of $Z_3$ is nonzero ($u_3^{(j)}\ne0$).
Then the Nahm equation tells that the $j$-th block of $Z_I$ is
independent of $y$. Similar argument holds with the role of $Z_I$ and
$Z_A$ exchanged.

An exceptional case is when $Z_A=0$. Then, the algebraic equations are
all satisfied without giving rise to any condition on $Z_I$, and the
scalars $Z_I$ can depend on $y$ in a non-trivial manner. As an obvious
generalization, if $Z_A$ takes the block-diagonal form (\ref{ZAbv}) and
one of the blocks is null, namely $u^{(j)}_A=0$ for some $j$, then $Z_I$
can show nontrivial $y$-dependence within the $j$-th diagonal
block. Again, the same holds with the role of $Z_A$ and $Z_I$ exchanged.

M5-branes can be regarded as initial conditions for the Nahm equation.
From the above analysis, one finds that the M5-branes behave like
D4-branes, i.e. M2-branes develop nontrivial Nahm pole at their end,
only when they are put on the $\mathbb Z_k$ orbifold singularity. If
they are away from the singularity, they behave more like NS5-branes
from the viewpoint of M2-branes ending on them. This is somewhat
puzzling, since the difference in the behavior of M5-branes persists
even in the case $k=1$ where the target space does not have any orbifold
singularity.

\section{Worldsheet theory on self-dual strings}\label{sec:worldsheet}

Let us now focus on the more specific situation of $N$ M2-branes
suspended between a pair of M5-branes (013456), one at $y=0$ and the
other at $y=L$. Their position in the $x_{7,8,9,\natural}$ direction is
related to the boundary condition $Z_A=u_A\cdot{\bf 1}_{N\times N}$. For
sufficiently small $L$ we expect to find a 2D theory with ${\cal N}=(4,2)$
supersymmetry on the M2-brane worldvolume.

\subsection{Derivation via dimensional reduction}\label{sec:ABJMslab}

We first derive the 2D theory by the standard dimensional reduction of
the ABJM model. In this approach, 2D Lagrangian is simply $L$ times the
3D Lagrangian with the $y$-dependence of fields dropped. The resulting
theory will be called {\it the ABJM slab} in this paper.

Let us look into each of the four terms in ${\cal L}$
(\ref{LABJM}). After the dimensional reduction, the Chern-Simons term
with the gluing condition $A_\mu=\tilde A_\mu$ becomes
\begin{equation}
 {\cal L}_\text{CS}~=~ \frac{k}{2\pi}\text{Tr}
 \big(\sigma F_{01}\big),\quad
 \sigma\equiv A_2-\tilde A_2\,.
\end{equation}
We note that $\sigma$ cannot be simply gauged away since, for arbitrary
$L$, the trace of the path-ordered exponential
\begin{equation}
 \text{Tr}\left[
 \text{P}\exp\left(i\int_0^L dy A_y\right)
 \text{P}\exp\left(i\int_L^0 dy\tilde A_y\right)
 \right]
\end{equation}
gives a gauge-invariant observable thanks to the gluing condition on
gauge fields. We also note that $\sigma$ is periodic, as $A_2=0$ and
$A_2=X$ are related by a large gauge transformation for any $X$ such
that $e^{iLX}=1\in U(N)$.

The matter kinetic term becomes
\begin{equation}
 {\cal L}_\text{mat}=\text{Tr}\Big(
  i\bar\Psi_I\gamma^\mu D_\mu\Psi^I
 +i\bar\Psi_A\gamma^\mu D_\mu\Psi^A
 -\bar u^Au_A\sigma^2
 -D_\mu\bar Z^ID^\mu Z_I
 -D_y\bar Z^ID_y Z_I
 \Big),
\label{Lmat2}
\end{equation}
where the spinors $\Psi^I$ have only anti-chiral component, and
$\Psi^A$ have only chiral component. The kinetic term for the fermions
can be rewritten in a more explicit form,
\begin{equation}
 -i\text{Tr}\Big(
   \bar\Psi_{I-}(D_0+D_1)\Psi^I_-
  +\bar\Psi_{A+}(D_0-D_1)\Psi^A_+
 \Big).
\end{equation}
The last term in (\ref{Lmat2}) cannot simply be dropped, since the
boundary condition on $Z_I$ is not of simple Neumann type.
The Yukawa and the potential terms become
\begin{eqnarray}
 {\cal L}_\text{yuk} &=& \frac{4\pi i}k\text{Tr}\Big(
  u_A\Psi^A[\bar Z^I,\bar\Psi_I]
 +\epsilon^{AB}u_A\bar\Psi_B
  \epsilon^{IJ}[Z_I,\bar\Psi_J]
 \nonumber \\ && \hskip10mm
 -\bar u^A\bar\Psi_A[Z_I,\Psi^I]
 -\epsilon_{AB}\bar u^A\Psi^B
  \epsilon_{IJ}[\bar Z^I,\Psi^J]
 \Big),
 \nonumber \\
 {\cal L}_\text{pot} &=&
 \frac{4\pi^2}{k^2}\bar u^Au_A\,
 \text{Tr}\Big(
   [Z_I,Z_J][\bar Z^I,\bar Z^J]
 - [Z_I,\bar Z^J][Z_J,\bar Z^I]\Big)
 \nonumber \\ &&
 -\frac{4\pi^2}{k^2}{\rm Tr}\Big(
  (\bar Z^JZ_J\bar Z^I-\bar Z^IZ_J\bar Z^J)
  (Z_I\bar Z^KZ_K-Z_K\bar Z^KZ_I)\Big).
\end{eqnarray}

If one naively uses the second of the boundary condition (\ref{bcsc}),
the last terms in ${\cal L}_\text{mat}$ and ${\cal L}_\text{pot}$ add
up. However, if the boundary contribution of (\ref{Lbd}) is added in the
form of a total $y$-derivative, the sum of the three can be rewritten
into the following form
\begin{eqnarray}
\lefteqn{
 -\text{Tr}\Big(
  D_y\bar Z^ID_yZ_I
 +\frac{4\pi^2}{k^2}
  (\bar Z^JZ_J\bar Z^I-\bar Z^IZ_J\bar Z^J)
  (Z_I\bar Z^KZ_K-Z_K\bar Z^KZ_I)
  \Big)
}
\nonumber \\
\lefteqn{
 +\frac{\pi}k\partial_y\text{Tr}\Big(
  \bar Z^IZ_I\bar Z^JZ_J-Z_I\bar Z^IZ_J\bar Z^J \Big)
}
\nonumber \\
 &=&
 -\text{Tr}\Big(
  D_y\bar Z^I-\frac{2\pi}k (\bar Z^JZ_J\bar Z^I-\bar Z^IZ_J\bar Z^J)\Big)
  \Big(D_yZ_I-\frac{2\pi}k (Z_I\bar Z^KZ_K-Z_K\bar Z^KZ_I)
  \Big),
\end{eqnarray}
which vanishes under the boundary condition (\ref{bcsc}).
It is interesting to notice here that, in our action integral
(\ref{Lbd}) for the M2-M5 system, the first (bulk) term is actually not
proportional to $L$ as is naively expected, but has an $L$-independent
piece. This piece is canceled by the boundary term, and the total
action integral is proportional to $L$.

With this cancellation understood, the remaining terms form a 2D
supersymmetric system. The Lagrangian for generic nonzero $u_A$ is given
by
\begin{eqnarray}
 \frac1L\cdot{\cal L} &=& \text{Tr}\Big(
  \frac{k}{2\pi}\sigma F_{01}
 +i\bar\Psi_I\gamma^\mu D_\mu\Psi^I
 +i\bar\Psi_A\gamma^\mu D_\mu\Psi^A
 -D_\mu\bar Z^ID^\mu Z_I-u^2\sigma^2
 \Big)
 \nonumber \\ && +\frac{4\pi u}k\text{Tr}\Big(
  \Psi^{\dot 3}[\bar Z^I,\bar\Psi_I]
 -\bar\Psi_{\dot 4}\epsilon^{IJ}[Z_I,\bar\Psi_J]
 +\bar\Psi_{\dot 3}[Z_I,\Psi^I]
 +\Psi^{\dot 4}\epsilon_{IJ}[\bar Z^I,\Psi^J]
 \Big)
 \nonumber \\ &&
 -\frac{4\pi^2u^2}{k^2}\text{Tr}
  \Big( [Z_I,\bar Z^J][Z_J,\bar Z^I]+[Z_I,Z_J][\bar Z^J,\bar Z^I] \Big),
\label{44qYM}
\end{eqnarray}
where
\begin{equation}
 \Psi^{\dot 3}=\frac iu u_A\Psi^A,\quad
 \Psi^{\dot 4}=-\frac iu \epsilon_{AB}\bar u^A\Psi^B.
 \quad
 \Big(u\equiv \sqrt{u_A\bar u^A}\Big)
\label{uPsi}
\end{equation}

We expected that the dimensionally reduced Lagrangian preserves
${\cal N}=(4,2)$ supersymmetry but the nonzero $u_A$ breaks the
$SU(2)\times SU(2)\times U(1)$ R-symmetry down to an $SU(2)$. But it
turns out that the system actually has ${\cal N}=(4,4)$ supersymmetry,
and the R-symmetry enhances to $SU(2)^3$. Indeed if one introduces the
notation
\begin{eqnarray}
 \text{scalar}&:&
 Z^{I\dot I}\, \equiv (Z^{I\dot1}\,,\,Z^{I\dot2}\,) \equiv
 (\epsilon^{IJ}Z_J,\bar Z^I),
 \nonumber \\
 \text{chiral spinor}&:&
 \Psi^{\dot I\dot A} \equiv (\Psi^{\dot1\dot A},\Psi^{\dot2\dot A}) \equiv
 (\Psi^{\dot A},-\epsilon^{\dot A\dot B}\bar\Psi_{\dot B}),
 \nonumber \\
 \text{anti-chiral spinor}&:&
 \Psi^{I\dot A} \equiv (\Psi^{I\dot3}\,,\Psi^{I\dot4}\,) \equiv
 (\Psi^I\,,-\epsilon^{IJ}\bar\Psi_J),
\label{bispinors}
\end{eqnarray}
one can show that the above Lagrangian is invariant under three copies
of $SU(2)$'s acting on the indices $I,\dot I$ and $\dot A$
respectively. The Lagrangian is also invariant under ${\cal N}=(4,4)$
supersymmetry transformation,
\begin{eqnarray}
 \delta Z^{I\dot I} &=&
  \xi_{~\dot A}^{\dot I}\Psi^{I\dot A}
 +\xi_{~\dot A}^{I}\Psi^{\dot I\dot A},
 \nonumber \\
 \delta\Psi^{\dot I\dot A} &=&
 -i\gamma^\mu\xi^{~\dot A}_{I}D_\mu Z^{I\dot I}
 +i\xi^{\dot I\dot A}\sigma u
 -\frac{2\pi u}k\xi^{~\dot A}_{\dot J}\epsilon_{IJ}
  [Z^{I\dot I},Z^{J\dot J}],
 \nonumber \\
 \delta\Psi^{I\dot A} &=&
 -i\gamma^\mu\xi^{~\dot A}_{\dot I}D_\mu Z^{I\dot I}
 -i\xi^{I\dot A}\sigma u
 -\frac{2\pi u}k\xi^{~\dot A}_{J}\epsilon_{\dot I\dot J}
 [Z^{I\dot I},Z^{J\dot J}],
 \nonumber \\
 \delta A_\mu &=& -\frac{2\pi u}k\Big(
  \xi_{I\dot A}\gamma_\mu\Psi^{I\dot A}
 +\xi_{\dot I\dot A}\gamma_\mu\Psi^{\dot I\dot A}\Big),
 \nonumber \\
 \delta \sigma &=& -\frac{2\pi i}k\Big(
  \xi_{\dot I\dot A}[Z^{I\dot I},\Psi^{~\dot A}_{I}]
 -\xi_{I\dot A}[Z^{I\dot I},\Psi^{~\dot A}_{\dot I}] \Big),
\end{eqnarray}
of which the ${\cal N}=(4,2)$ part can be obtained from the reduction of
3D unbroken supersymmetry. Here the supersymmetry parameter
$\xi^{\dot I\dot A}$ is chiral and $\xi^{I\dot A}$ is anti-chiral.

Most importantly, the Lagrangian for the ABJM slab looks like a
$q$-deformed version of ${\cal N}=(4,4)$ super Yang-Mills theory.
Indeed, if one forgets about the periodicity of $\sigma$ and naively
integrates it out, we obtain the kinetic term for the gauge field
\begin{equation}
 {\cal L} = -\frac1{2g^2}\text{Tr}(F_{\mu\nu}F^{\mu\nu})+\cdots,\quad
 g^2 =\frac{16\pi^2 u^2}{k^2L}.
\label{gym2}
\end{equation}
A natural question arises at this point: how is the $\sigma^2$ term in
the Lagrangian consistent with the periodicity of $\sigma$?

In fact, the Lagrangian including the auxiliary scalar $\sigma$ have
been used frequently in the study of 2D pure Yang-Mills theory.
In particular, \cite{Aganagic:2004js} proposed the so-called
$q$-deformed Yang-Mills theory by making $\sigma$ a periodic variable.
The Lagrangian for the ABJM slab contains basically the same Yang-Mills
term and periodic $\sigma$, but in its derivation the $\sigma^2$ term arose
in a rather strange way through the boundary value of charged matter
scalars. Therefore, the ABJM slab may well answer to the above question
differently from the $q$-deformed Yang-Mils theory studied in
\cite{Aganagic:2004js}. We discuss this point in detail in the next
subsection.

In the special case $u_A=0$, many terms in the Lagrangian disappear,
and one is left with a topological gauge theory coupled to adjoint matters,
\begin{equation}
 \frac1L\cdot{\cal L} ~=~
 \text{Tr}\Big(\frac{k}{2\pi}\sigma F_{01}
 +i\bar\Psi_I\gamma^\mu D_\mu\Psi^I
 +i\bar\Psi_A\gamma^\mu D_\mu\Psi^A
 -D_\mu\bar Z^ID^\mu Z_I \Big)\,.
\label{44TYM}
\end{equation}
The chiral spinors $\Psi^{\dot A}$ (\ref{uPsi}) become ill-defined, but
one can instead introduce
\begin{equation}
 \Psi^{\dot IA} \equiv(\Psi^{\dot 1A},\Psi^{\dot 2A})\equiv
 (\Psi^A,-\epsilon^{AB}\bar\Psi_B)
\end{equation}
and rewrite the Lagrangian using the scalars $Z^{I\dot I}$, anti-chiral
spinors $\Psi^{I\dot A}$ defined in (\ref{bispinors}) and the above chiral
spinors $\Psi^{\dot IA}$. One can then explicitly see that the
R-symmetry is enhanced to $SU(2)^4$, and the supersymmetry is generated
by chiral spinors $\xi^{\dot I\dot A}_+$ and anti-chiral spinors $\xi^{IA}_-$.

The moduli space of vacua of the ABJM slab for generic $u>0$ is that of
${\cal N}=(4,4)$ $U(N)$ super Yang-Mills theory, so it is the $N$-th
symmetric product of $\mathbb R^4$. It is parametrized by a pair of
$N\times N$ matrices $Z_I$ which commute with themselves and their
Hermite conjugates. Since the effect of $\mathbb Z_k$ orbifold is to
identify the system at different values of the coupling $u_A$, we do not
need to care about its action on the fields. For $u=0$, on the other
hand, the scalar potential vanishes and the relation between the moduli
space and the position of M2-branes becomes unclear. Moreover one has to
take care of the $\mathbb Z_k$ orbifold acting on the fields in this case.

\subsection{$q$-deformed Yang-Mills revisited}\label{sec:qYM}

Here we would like to compare some aspects of the 2D SUSY gauge theories
obtained above with those of 2D pure Yang-Mills theory and its $q$-deformation,
paying special attention to the periodicity of the scalar field. Let us
first review some of the exact analysis of 2D Yang-Mills and 3D
Chern-Simons theories.

It is a special feature of two dimensions that the Yang-Mills action
depends only on the volume and not on any more detail of the metric of
the surface. A nice way to see this feature is to regard it as a
deformation of a topological field theory. Let $\Sigma_h$ be a Riemann
surface of genus $h$. The Euclidean action on $\Sigma_h$ is
\begin{equation}
 S_\text{YM} ~=~ \frac1{2g^2}\int_{\Sigma_h}\text{Tr}
 \big(F\wedge\ast F\big)
 ~\simeq~ \int_{\Sigma_h}\text{Tr}
 \Big(i\phi F + \frac{g^2}2\omega\phi^2\Big).
\end{equation}
Here $\omega$ is the normalized volume form of $\Sigma_h$ satisfying
$\int_{\Sigma_h}\omega=1$. The theory at $g=0$ does not depend on the
metric at all, so it is a topological field theory. Even after $g$ is
turned on, the theory is invariant under area-preserving
diffeomorphisms. Thanks to this property the 2D Yang-Mills theory is
exactly solved; the partition function on Riemann surfaces of arbitrary
genus has been worked out by various techniques, for example the idea of
lattice gauge theory was applied in \cite{Witten:1991we}, and in
\cite{Witten:1992xu} more detail of the theory was studied using
non-abelian localization.

Let us look more closely into the theory using the approach of
abelianization \cite{Blau:1993tv}. This approach is characterized by the
gauge choice in which $\phi$ is diagonal. To be more explicit, let us
denote a Cartan subalgebra of the gauge symmetry algebra by ${\bf t}$
and its complement by ${\bf k}$. We decompose various fields
accordingly, as $\phi=\phi^{\bf t}+\phi^{\bf k}$ etc. Our (partial)
gauge-fixing condition is therefore $\phi^{\bf k}=0$. The remaining
${\bf t}$-gauge invariance can be fixed in any convenient way; for
example the Lorentz gauge ${\rm d}\ast A^{\bf t}=0$.

The path integral of the system including ghosts was studied in
\cite{Blau:1993tv}. The strategy is to integrate first over all the
fields except $\phi^{\bf t}$, which in fact simply gives rise to a delta
function and a determinant. The delta function arises
from the integration of ${\bf t}$-valued fields, which requires $\phi^{\bf t}$
to be constant and take values on the weight lattice of the gauge group.
The Gaussian integral over ${\bf k}$-valued fields gives rise to the
determinant of $\phi$ acting on (${\bf k}$-components of) the ghosts
$c,\bar c$ and the gauge fields $A$,
\begin{equation}
 \frac{\text{Det}(\phi)|_{\bar c^{\bf k},c^{\bf k}}}
      {(\text{Det}(\phi)|_{A^{\bf k}})^{1/2}}
 ~=~ \Big(\prod_{\alpha>0}\alpha\cdot\phi\Big)^{2-2h}.
\end{equation}
Here the product is over all the positive roots $\alpha$. Identifying
points on the weight lattice with gauge group representations, one can
express the partition function on genus-$h$ surface as follows,
\begin{equation}
 Z_h~=~ \text{const}\cdot\sum_{\lambda:\text{reps.}}
 d(\lambda)^{2-2h}e^{-\frac12g^2C_2(\lambda)},
\label{ZYM}
\end{equation}
where $C_2(\lambda)=\lambda\cdot(\lambda+2\rho)$ is the second Casimir,
$\rho$ is the Weyl vector and $d(\lambda)$ is the dimension of the
representation $\lambda$.
\begin{equation}
 d(\lambda)\equiv
  \prod_{\alpha>0}\frac{\alpha\cdot(\lambda+\rho)}{\alpha\cdot\rho}.
\end{equation}

The $q$-deformation introduces a periodicity to $\phi$ by regarding it
not simply as an adjoint scalar but a component of the gauge field along
a compact direction. It was introduced in \cite{Aganagic:2004js} in
the study of D4-branes in the topological string theory on a non-compact
Calabi-Yau manifold. The target is given by two line bundles $L_1\oplus L_2$
fibred over a Riemann surface $\Sigma_h$, satisfying the Calabi-Yau condition
\begin{equation}
 \text{deg}(L_1)+\text{deg}(L_2)=2h-2.
\end{equation}
The D4-branes support a topological gauge theory with the Lagrangian
$-\frac1{2g_s}\text{Tr}F\wedge F$. For $N$ D4-branes wrapping the
four-dimensional non-compact submanifold $L_2\to\Sigma_h$, the theory
becomes the $U(N)$ Chern-Simons theory of level $k=2\pi i/g_s$ on its
boundary, which is a circle bundle over $\Sigma_h$ of monopole degree
$p=\text{deg}(L_2)$. Let us denote such a 3-manifold by $M_{h,p}$. It was
argued in \cite{Aganagic:2004js} and explained in full detail in
\cite{Blau:2006gh} that the worldvolume gauge theory can be further
reduced to a 2D $q$-deformed Yang-Mills theory on $\Sigma_h$.

The topological structure of $M_{h,p}$ is characterized by a 1-form
$\kappa$ satisfying ${\rm d}\kappa=2\pi p\omega$, with $\omega$ the
normalized volume form on $\Sigma_h$ as before. Using the coordinate
$\theta\sim\theta+2\pi$ for the circle fiber and $x^1,x^2$ for the base
$\Sigma_h$, it can be expressed as $\kappa={\rm d}\theta+a_i{\rm d}x^i$.
The exterior derivative on $M_{h,p}$ can be decomposed as
${\rm d}=\kappa\partial_\theta +{\rm d}_H$, and
${\rm d}_H={\rm d}x^i(\partial_i-a_i\partial_\theta)$ can be used to
define parallel transport along the base $\Sigma_h$.
Using the decomposition of the gauge field $A_{(3)}=\phi\kappa+A$ into
the base and fiber directions, the Chern-Simons action can be written as
\begin{eqnarray}
 S &=& \frac {ik}{4\pi}\int_{M_{h,p}}
 \text{Tr}\Big(A_{(3)}{\rm d}A_{(3)}-\frac{2i}3A_{(3)}^3\Big)
 \nonumber \\ &=&
 \frac {ik}{4\pi}\int_{M_{h,p}}
 \text{Tr}\Big(2\pi p\phi^2\kappa\wedge\omega
  +2\phi\kappa\wedge {\rm d}A
  +\kappa\wedge(\partial_\theta A-i[\phi,A])\wedge A
 \Big).
\end{eqnarray}
Here the overall $i$ is because the Euclidean path integral weight is
$e^{-S}$ in our convention.

All the fields are periodic function of $\theta$ and therefore can be
decomposed into Fourier modes. The $n$-th Fourier modes couple to the
gauge field $A+na$ on $\Sigma_h$, so they are the sections of
${\cal O}(np)$ on $\Sigma_h$ besides carrying the gauge charge. The
Chern-Simons theory would reduce to the 2D Yang-Mills theory with gauge
coupling $\sim p^{1/2}$ if one discarded all the non-zero
modes. But in fact, no matter how small the radius of the fiber circle
becomes, one can never simply discard the nonzero modes, as the
Chern-Simons theory is a topological field theory.

For the computation of partition function, it is most convenient to work
in the gauge $\phi^{\bf k}=\partial_\theta\phi^{\bf t}=0$. As before,
the strategy is to path-integrate over all the fields except $\phi^{\bf t}$.
Those fields are further divided into two groups. The first consists
of the ${\bf t}$-valued $\theta$-independent modes; integration over
them yields a delta function requiring $\phi$ to take quantized constant
values. The second group contains all the remaining modes; integrating
over them gives rise to the familiar shift of the level by the dual
Coxeter number $k\to \hat k\equiv k+h^\vee$, as well as the determinant
\begin{equation}
 \Big(\prod_{\alpha>0}\sin(\pi\alpha\cdot\phi)\Big)^{2-2h}.
\end{equation}
The determinant has the appropriate periodicity in $\phi$ thanks to all the
Fourier modes having been taken into account. The delta function constraint
now requires that $\hat k\cdot\phi^{\bf t}$ be on the weight lattice.
Thus the partition function of $q$-deformed Yang-Mills theory is given
by the same formula (\ref{ZYM}), with the dimension of the representations
replaced by its $q$-analogue,
\begin{equation}
 d_q(\lambda)\equiv\prod_{\alpha>0}
 \frac{[\alpha\cdot(\lambda+\rho)]_q}{[\alpha\cdot\rho]_q},\quad
 [x]_q\equiv\frac{q^{x/2}-q^{-x/2}}{q^{1/2}-q^{-1/2}}\,.
\end{equation}
The deformation parameter is $q\equiv \exp(2\pi i/\hat k)$.

Note that for integer $k$ the number of representations contributing to
the partition function becomes finite. The 2D Yang-Mills coupling also
depends on the renormalized Chern-Simons coupling as $g^2=i\hat kp$. The
pure imaginary and quantized value of $g^2$ is essential for the
periodicity of $\phi$.

\vskip3mm

Let us now turn to the model of our interest. If one forgets about the
matter fields, the gauge sector is $U(N)\times U(N)$ Chern-Simons theory
on the slab $\mathbb R^{1,1}\times (\text{interval})$, with the
identification of the two $U(N)$ gauge fields at $y=0$ and $y=L$. This
is effectively the same as the $U(N)$ Chern-Simons theory on
$\mathbb R^{1,1}\times S^1$, corresponding to the case $p=0$ in the
previous paragraph. Therefore, in the absence of matters the theory
would not reduce to 2D topological Yang-Mills theory in the small $L$
limit. Actually it is known that for $p=0$ a different 2D gauge theory,
called $G/G$ gauged WZNW model with $G=U(N)$, gives a precise
description of the 3D system \cite{Blau:1993hj}.

The term $u^2\sigma^2$ in (\ref{44qYM}) arises from the nonzero boundary
value of matter scalar fields obeying Dirichlet boundary condition.
Unlike the Chern-Simons theory on $M_{h,p}$, this term (in Euclidean
action) is real, and there is no reason that its coefficient is
quantized. So we need a different argument to resolve the contradiction
that such a term is present in a system of periodic $\sigma$.

To make the argument simple and concrete, we limit our discussion here
to the simplest example of Euclidean $U(1)\times U(1)$ ABJM model on
$S^2\times\text{(interval)}$, though it is not fully clear how the
following argument extends to the more general cases. As is shown in the
appendix, if the $S^2$ is round, one can introduce auxiliary fields to
make the 2D off-shell supersymmetry manifest.

If one denotes the average and difference of the two $U(1)$ gauge fields
by $B_m$ and $C_m$, the Lagrangian becomes
\begin{equation}
 {\cal L}~=~ \frac{ik}{4\pi}\varepsilon^{mnp}B_m\partial_nC_p
 +D_m\bar Z^aD^mZ_a+(\text{fermions}),
\end{equation}
where the matter covariant derivative is defined as
$D_mZ_a=(\partial_m-iC_m)Z_a$, etc. The Yukawa and bosonic potential
terms all happen to vanish. The $B_m$ equation of motion requires $C_m$
to be flat. So the path integral reduces to that over a flat gauge field
$C_m$ and some free matter fields coupled to it.

Among the path integration variables is a real constant field
$\sigma=C_y$ which has periodicity $\sigma\sim\sigma+2\pi/L$. The
periodicity arises from the large gauge transformations
\begin{equation}
 \sigma'=\sigma+\frac{2\pi n}L,\quad
 Z'_a=Z_a \exp\Big(\frac{2\pi iny}L\Big).\quad(n\in\mathbb Z)
\label{lgt}
\end{equation}
On the other hand, when deriving the dimensionally reduced Lagrangian
(\ref{44qYM}), we assumed $Z_A=u_A=(\text{constant})$ and neglected all
the Kaluza-Klein modes, which clearly breaks the large gauge invariance.
To restore the invariance, we need to take account of all the matter
Kaluza-Klein modes. The scalars $Z_I$ and $Z_A$ obey Neumann and
Dirichlet boundary conditions respectively, so for constant $C_y=\sigma$
they are naturally expanded into the Fourier modes
\begin{eqnarray}
 Z_I (x^\mu,y)&=&\sum_{n\ge0}Z_{I(n)}(x^\mu)\cdot
 e^{i\sigma y}\cos\frac{\pi n y}L,
 \nonumber \\
 Z_A (x^\mu,y)&=&u_A+\sum_{n>0}Z_{A(n)}(x^\mu)\cdot
 e^{i\sigma y}\sin\frac{\pi n y}L.
\label{Zfourier}
\end{eqnarray}
Now if the large gauge transformation (\ref{lgt}) is applied to $Z_A$,
its constant piece $u_A$ is transformed into an oscillating function,
but the change can be absorbed by appropriate shifts of the mode
variables $Z_{A(n)}$.

In order to see how the term $L u^2\sigma^2$ in the Lagrangian gets
modified by the Kaluza-Klein modes, we integrate out the free matter
fields and see how the result depends on $\sigma$. We notice that,
assuming $\sigma$ is constant, the $\sigma$-dependence in the Lagrangian
and the Fourier decomposition (\ref{Zfourier}) can be eliminated almost
completely by a gauge transformation from $C_y=\sigma$ to $C_y=0$. This
transformation is a symmetry of the system though not in the group of
gauge equivalence. After that the only $\sigma$-dependence remains in
the term $D_y\bar Z^AD_yZ_A$ in the Lagrangian. By substituting
(\ref{Zfourier}) into it, we obtain
\begin{equation}
\int_0^L{\rm d}y D_y\bar Z^AD_yZ_A =
 \frac1L\sum_{n>0}\Big(
   \frac{n^2\pi^2}2\bar Z^A_{(n)}Z_{A(n)}
   +\bar Z_{A(n)}c_{A(n)}+ c_{A(n)}^\ast
   Z_{A(n)}\Big)+Lu^2\sigma^2,
\end{equation}
with
\begin{equation}
 c_{A(n)} = u_A
 \frac{L^2\sigma^2 n\pi (e^{-iL\sigma+in\pi}-1)}
      {L^2\sigma^2-n^2\pi^2} .
\end{equation}
The Gaussian integral over the mode variables $Z_{A(n)}, Z_{I(n)}$ and
their superpartners does not yield $\sigma$-dependent determinant, but
there remains a $\sigma$-dependent classical Lagrangian
\begin{equation}
 {\cal L}_\text{cl}= Lu^2\sigma^2
 -\sum_{n>0}\frac{2{c^\ast}_{\!\!\!A(n)}c_{A(n)}}{Ln^2\pi^2}
 = \frac{4u^2}L\sin^2\frac{\sigma L}2\,.
\label{Lcl}
\end{equation}
This is the desired periodic function which approaches $Lu^2\sigma^2$
for small $\sigma L$.

The Lagrangian (\ref{Lcl}) can be well approximated by the quadratic
function and $\sigma$ can be integrated out if the saddle of the
Gaussian $\sigma$-integration is sufficiently close to $\sigma=0$ and
the width is narrow. One finds from (\ref{44qYM}) that the location of
the saddle and the width (in terms of the variable of unit periodicity
$L\sigma$) are
\begin{equation}
 L\sigma = \frac{kL}{4\pi u^2}F_{01} = \frac{4\pi}{kg^2}F_{01},\quad
 \langle(L\sigma)^2\rangle = \frac{8\pi^2}{k^2g^2}\,.\quad
 \Big(g^2\equiv\frac{16\pi^2u^2}{k^2L}\Big)
\label{sadsigma}
\end{equation}
Therefore the periodicity becomes unimportant in the limit $L\to 0$.

More careful argument would take into account that the average value of
$F_{01}$ may also depend on $L$. It would be reasonable to guess such an
effect from the dependence of $\langle F_{01}\rangle$ of 2D
${\cal N}=(4,4)$ super Yang-Mills theory on the gauge coupling
$g^2\sim u^2/L$. $\langle F_{01}\rangle$ would therefore stay constant
in the limit of small $L$ if we send $u\to 0$ at the same time so that
$g^2$ is kept fixed. In view of (\ref{sadsigma}), again we believe that
in the limit $L\to 0$ with $u$ fixed the periodicity will become
unimportant.

\subsection{Derivation via brane construction}\label{sec:IIAbrane}

There is another candidate for the worldsheet theory of self-dual
strings which was obtained via a dual type IIA brane construction
\cite{Haghighat:2013tka}. We will hereafter call it {\it the IIA brane
model}. This model can be compared with the ABJM slab at least for
$k=1$, in which case the transverse geometry is simply
$\mathbb R^8=\mathbb R^4_{(3456)}\times\mathbb R^4_{(789\natural)}$.
Here we briefly review the construction of the model.

Let us replace the $\mathbb R^4_{(789\natural)}$ by a Taub-NUT space,
that is a circle fibration over $\mathbb R^3$ such that the radius of
the circle asymptotes to a constant $R$ at infinity but shrinks at the
origin of $\mathbb R^3$. It becomes the flat $\mathbb R^4$ in the limit
$R\to\infty$, so the replacement of the background does not affect the
values of $R$-independent observables.

In the limit of small $R$ we move to the weakly coupled type IIA
superstring theory with a single D6-brane (0123456), and the M2-M5
system turns into the D2-branes (012) suspended between parallel
NS5-branes (013456). The $N$ D2-branes suspended between two NS5-branes
give rise to the $U(N)$ ${\cal N}=(4,4)$ super Yang-Mills theory on the
worldvolume, but the
D6-brane breaks half of the supersymmetry and introduces additional
fundamental matters. Although the supersymmetry of the D2-D6-NS5 system
is half of the M2-M5 system, their worldvolume theory was shown to
reproduce the elliptic genus of self-dual strings evaluated using
refined topological vertex formalism \cite{Haghighat:2013gba,Haghighat:2013tka}.

Let us choose the orientation of the IIA branes so that the unbroken
supersymmetry is given by the (10D 32-component spinor) supercharge
$Q$ satisfying
\begin{equation}
 Q=\Gamma^{012}Q=\Gamma^{013456}Q=-\Gamma^{0123456}Q,
\end{equation}
and corresponds to ${\cal N}=(0,4)$ in two dimensions. This is in order
for the elliptic genus to be a holomorphic function of the modular
parameter of the torus in the convention of
\cite{Benini:2013nda,Benini:2013xpa}. The brane configuration is
summarized in the table below.

\begin{center}
 \begin{tabular}{|c|cccccccccc|}
 \hline
   & 0 & 1 & 2 & 3 & 4 & 5 & 6 & 7 & 8 & 9 \\
 \hline
  D6  &$-$&$-$&$-$&$-$&$-$&$-$&$-$&  &  &  \\
 \hline
  D2  &$-$&$-$&$\dashv$&  &  &  &  &  &  &  \\
 \hline
  NS5 &$-$&$-$& &$-$&$-$&$-$&$-$&  &  &  \\
 \hline
 \end{tabular}
\end{center}

To understand the structure of the 2D theory, it is helpful to consider
first the parallel D2-D6 system without NS5-branes, which gives rise to
the 3D ${\cal N}=4$ $U(N)$ gauge theory with one adjoint and one fundamental
hypermultiplets. The vector multiplet contains a gauge field (one of its
components becomes scalar upon dimensional reduction) and three scalars
describing the motion of the D2-branes in the $x_{7,8,9}$
directions. The scalars in the adjoint hypermultiplet are for the motion
in the $x_{3,4,5,6}$ direction, and the fundamental hypermultiplet
arises from D2-D6 strings. We reduce this theory to two dimensions,
and denote the fields in the 2D ${\cal N}=(4,4)$ theory as follows.
\TABLE{
\def\arraystretch{1.2}
\begin{tabular}{|r|cccc|}
 \hline
  multiplet &
  scalar &
  \multicolumn{2}{c}{spinor} &
  vector \\ \hline
  vector&
  $Y^{A\dot A}$, &
  $\lambda^{\dot I\dot A}_+$, &
  $\lambda^{\dot IA}_-$, &
  $A_\mu$, \\
  adjoint hyper &
  $Z^{I\dot I}$, &
  $\Psi^{IA}_+$, &
  $\Psi^{I\dot A}_-$, &
  \\
  fundamental hyper &
  $q^{\dot I}$, &
  $\psi^A_+$, &
  $\psi^{\dot A}_-$, &
  \\ \hline
\end{tabular}
\def\arraystretch{1.0}
\label{table:44fields}
\caption{Fields in the 2D ${\cal N}=(4,4)$ theory.}
}

The 2D theory has $SU(2)^3$ R-symmetry and an $SU(2)$ flavor symmetry
acting on the adjoint hypermultiplet. The indices $I,\dot I,A,\dot A$
are for the doublets under the symmetry $SU(2)_1,\cdots,SU(2)_4$, where
$SU(2)_1$ is the flavor symmetry and the other three are R-symmetries.
The ${\cal N}=(4,4)$ supersymmetry is parametrized by the spinors
$\xi_+^{\dot I\dot A}$ and $\xi_-^{\dot IA}$.

The NS5-branes provide the Dirichlet boundary condition on
$Y^{A\dot A},\lambda_-^{\dot IA}$ and $\Psi_+^{IA}$, but other
fields obey Neumann boundary condition. They also break the
supersymmetry parametrized by $\xi^{\dot IA}_-$.

In this construction, one can choose freely where to put the M2-M5
system in the transverse Taub-NUT geometry. This corresponds to the
choice of boundary conditions
$Y^{A\dot A}=u^{A\dot A}\cdot{\bf 1}_{(N\times N)}$. For generic nonzero
$u^{A\dot A}$ the fundamental hypermultiplet fields become all
massive. Also, the $SU(2)_3\times SU(2)_4$ symmetry is broken to
a diagonal subgroup. In the limit of large $u^{A\dot A}$ the fundamental
hypermultiplet gets frozen, and the resulting system of adjoint fields
is actually the ${\cal N}=(4,4)$ super Yang-Mills theory with R-symmetry
$SU(2)_1\times SU(2)_2\times SU(2)_\text{diag}$, with the supersymmetry
parametrized by $\xi^{\dot I\dot A}_+$ and $\xi^{I\dot A}_-$.
This theory is in agreement with the ABJM slab when the M2-M5
system is away from the origin of $\mathbb R^4_{(789\natural)}$.

On the other hand, for $u^{A\dot A}=0$ the theory has only
${\cal N}=(0,4)$ supersymmetry but an enhanced global symmetry
$SU(2)^4$, since $SU(2)_3$ and $SU(2)_4$ become independent.
As compared to the ABJM slab at $u_A=0$, the global symmetry matches but
the supersymmetry does not agree.

Thus the two derivations of the worldsheet theory of self-dual strings, 
using ABJM and IIA brane models, led to considerably different results.
Part of the difference, for example the mismatch of
symmetry and SUSY, is because we broke the symmetry explicitly on the
IIA side by replacing the background $\mathbb R^4$ by a Taub-NUT.
If the two derivations are both valid, then there should be a smooth
interpolation of the two models corresponding to changing the asymptotic
radius $R$ of the circle fiber of the Taub-NUT.
One can also check the validity by evaluating $R$-independent physical observables
in the two descriptions and making comparison.

\subsection{Quiver models}\label{sec:IIAquiver}

Let us briefly comment on the case with more than two M5-branes. Suppose
there are $K$ M5-branes at the origin of $\mathbb R^4_{(789\natural)}$
mutually separated in the $y$-direction. The $i$-th M5-brane is at
$y=y_i$, where $y_1<y_2<\cdots<y_K$. In addition, we have $N_i$
M2-branes suspended in the $i$-th interval $[y_i,y_{i+1}]$. The
corresponding IIA brane system is described in Figure \ref{fig:IIAquiver}.

\FIGURE{
\setlength\unitlength{25pt}
\begin{picture}(15,7)(0,-0.3)
\put(0,0){\framebox(15,6)}
\put( 0.1,5.6){\footnotesize D6}
\linethickness{1.6pt}
\put( 1.2,0.0){\line(0,1){6}}
\put( 6.0,0.0){\line(0,1){6}}
\put( 9.0,0.0){\line(0,1){6}}
\put(12.5,0.0){\line(0,1){6}}
\put( 0.9,6.2){\footnotesize\bf NS5}\put( 0.9,-0.3){\footnotesize$y_{i-1}$}
\put( 5.7,6.2){\footnotesize\bf NS5}\put( 5.8,-0.3){\footnotesize$y_i$}
\put( 8.7,6.2){\footnotesize\bf NS5}\put( 8.7,-0.3){\footnotesize$y_{i+1}$}
\put(12.2,6.2){\footnotesize\bf NS5}\put(12.2,-0.3){\footnotesize$y_{i+2}$}
\linethickness{0.6pt}
\put( 1.2,3.5){\line(1,0){4.8}}
\put( 1.2,3.6){\line(1,0){4.8}}
\put( 1.2,3.7){\line(1,0){4.8}}
\put( 2.6,3.1){\footnotesize\sf D2$(N_{i-1})$}
\put( 6.0,3.0){\line(1,0){3.0}}
\put( 6.0,3.1){\line(1,0){3.0}}
\put( 6.0,3.2){\line(1,0){3.0}}
\put( 6.0,3.3){\line(1,0){3.0}}
\put( 7.0,3.5){\footnotesize\sf D2$(N_i)$}
\put( 9.0,3.7){\line(1,0){3.5}}
\put( 9.0,3.8){\line(1,0){3.5}}
\put(10.2,3.3){\footnotesize\sf D2$(N_{i+1})$}
\linethickness{0.5pt}
\put( 8.5,3.0){\circle*{0.1}}
\put(10.0,1.8){\circle{0.1}}
\qbezier( 8.5,3.0)( 9.1,3.5)(9.99,1.86)
\put( 9.5,2.64){\vector(1,-1){0.01}}
\put(10.1,2.0){\footnotesize $\psi_{+(i)}$}
\put( 7.8,3.1){\circle*{0.1}}
\put( 7.2,1.8){\circle{0.1}}
\qbezier( 7.8,3.1)( 7.5,3.6)(7.21,1.86)
\put( 7.25,2.1){\vector(-1,-4){0.01}}
\put( 6.7,1.4){\footnotesize $q^{\dot I}_{(i)},\psi^{\dot A}_{-(i)}$}
\put( 6.5,3.0){\circle*{0.1}}
\put( 5.0,1.8){\circle{0.1}}
\qbezier( 6.5,3.0)( 5.9,3.5)(5.01,1.86)
\put( 5.6,2.73){\vector(-1,-1){0.1}}
\put( 4.2,2.1){\footnotesize $\tilde\psi_{+(i)}$}
\put( 8.3,3.3){\circle*{0.1}}
\put( 9.8,3.8){\circle*{0.1}}
\qbezier( 8.3,3.3)( 9.0,5.2)(9.79,3.86)
\put( 9.5,4.25){\vector(1,-1){0.1}}
\put(9.4,4.4){\footnotesize
 $Y^{\dot A}_{(i)},\lambda^{\dot I}_{-(i)},\Psi^I_{+(i)}$}
\put( 6.9,3.2){\circle*{0.1}}
\put( 5.3,3.7){\circle*{0.1}}
\qbezier( 6.9,3.2)( 6.0,5.2)(5.3,3.73)
\put( 5.49,4.04){\vector(-3,-4){0.1}}
\put(2.0,4.4){\footnotesize
 $\bar Y^{\dot A}_{(i-1)},\bar\lambda^{\dot I}_{-(i-1)},\bar\Psi^I_{+(i-1)}$}
\end{picture}
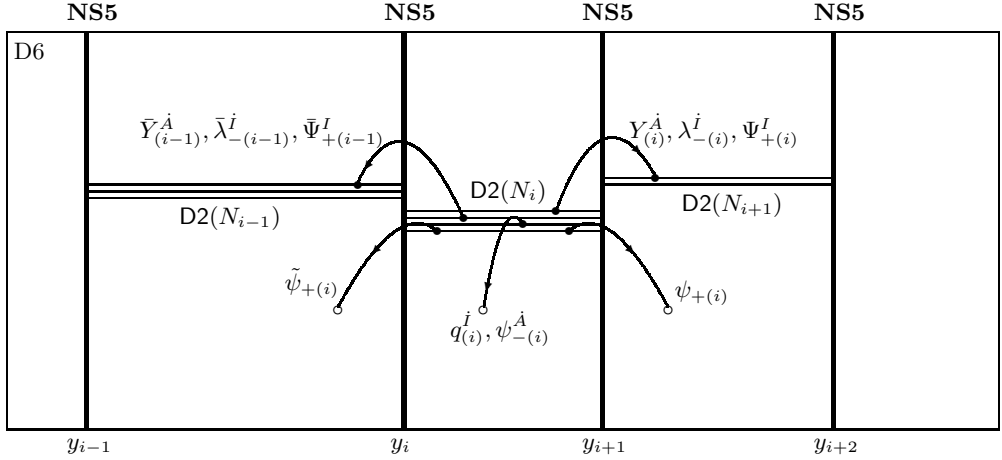
\caption{Type IIA brane system which gives rise to a quiver gauge
theory. The D2-D2 strings and D2-D6 strings in this figure correspond to
the bi-fundamental and fundamental matter fields, respectively.}
\label{fig:IIAquiver}
}

From the type IIA brane picture, we obtain the following 2D ${\cal N}=(0,4)$
SUSY gauge theory \cite{Haghighat:2013tka}. The gauge group is
$\otimes_{i=1}^{K-1}U(N_i)$, where the $U(N_i)$ arises from the
D2-branes in the $i$-th interval. Similarly to the model discussed in
Section \ref{sec:IIAbrane}, one has the adjoint and fundamental fields
for each $U(N_i)$,
\begin{equation}
\def\arraystretch{1.4}
\begin{array}{rcl}
 A_{\mu(i)},~Z^{I\dot I}_{(i)},~\Psi^{I\dot A}_{-(i)},~
 \lambda^{\dot I\dot A}_{+(i)}
 &~:~&
 \text{adjoint of }U(N_i) \\
 q^{\dot I}_{(i)},~\psi^{\dot A}_{-(i)},~\psi_{+(i)},~\tilde\psi_{+(i)}
 &~:~&
 {\bf N}_i\text{ of }U(N_i).
\end{array}
\def\arraystretch{1}
\end{equation}
Here we denoted by $\psi_{+(i)},\tilde\psi_{+(i)}$ the two components of
the fundamental chiral spinor $\psi_+^A$ in the Table \ref{table:44fields}.
In addition, there are bi-fundamental matter fields connecting the
neighboring gauge groups,
\begin{equation}
 Y^{\dot A}_{(i)},~\lambda^{\dot I}_{-(i)},~\Psi^I_{+(i)}~~:~~
 ({\bf N}_i,\overline{\bf N}_{i+1})\text{ of }U(N_i)\times U(N_{i+1}).
\end{equation}
In this quiver theory the $SU(2)_3$ is broken to $U(1)$. Under this
$U(1)$, the bi-fundamental fields
$Y^{\dot A}_{(i)},\lambda^{\dot I}_{-(i)},\Psi^I_{+(i)}$ carry the
charge $1/2$, the fundamental fields $(\psi_{+(i)},\tilde\psi_{+(i)})$
carry $(+1/2,-1/2)$, and all other fields are neutral. These fields
carrying nonzero $U(1)$ charge all correspond to open strings connecting
D-branes in the neighboring intervals, as shown in Figure
\ref{fig:IIAquiver}. The charge assignments to the fields and the
symmetry breaking can be most easily understood by considering a similar
system of branes with periodic identification of $x_2$ direction. After
T-dualizing along $x_2$ we obtain the system of D1, D5-branes in a
transverse $\mathbb Z_K$ orbifold, for which the standard construction
\cite{Douglas:1996sw} allows one to identify the worldvolume field theory.

To study the same system using ABJM model, one needs not only the
boundary condition but also the junction condition on fields for
M2-branes intersecting with M5-branes. With regard to this
aspect, our current understanding of the ABJM model is rather limited.
The similar system of D3-branes ending on or intersecting with D5-branes
was studied systematically in \cite{Gaiotto:2008sa} through the analysis
of Nahm equation and the moduli space of its solutions. There it was
shown that the physics at the D3-D5 intersection varies very much
depending on the number of D3-branes ending on the two sides of a
D5-brane. The system of M2 and M5-branes needs to be studied in a
similar manner.

\section{Elliptic genus}\label{sec:elliptic}

In this section we study the elliptic genus for the two 2D gauge
theories for self-dual strings. If the two theories are dual or
connected by some continuous deformation, their elliptic genera should
agree. Elliptic genus for general 2D ${\cal N}=(2,2)$ and ${\cal N}=(0,2)$
supersymmetric gauge theories has been studied recently in
\cite{Gadde:2013dda,Benini:2013nda,Benini:2013xpa}. On the other hand,
the elliptic genus of multiple self-dual strings has been derived in
\cite{Haghighat:2013gba} using topological vertex formalism.

Elliptic genus can be formulated for 2D theories with at least
${\cal N}=(0,2)$ supersymmetry as a partition function on a two-torus
with SUSY preserving boundary condition on fields, and is a holomorphic
function of the modulus $\tau$. If the theory has global symmetry that
commutes with ${\cal N}=(0,2)$ supersymmetry, one can gauge it by an
external flat gauge field $A_\mu$. Then the elliptic genus also depends
holomorphically on the fugacity parameter
$w\equiv \text{Im}\tau\cdot(A_1+iA_2)/2\pi i$.

\paragraph{Constructions of 2D SUSY theories.}

We begin by summarizing how 2D field theories with various
supersymmetry can be constructed from ${\cal N}=(0,2)$ supermultiplets.
Gauge theories with ${\cal N}=(0,2)$ supersymmetry generally consist of
three kinds of multiplets, namely vector multiplet
$(A_\mu,\lambda_+,\bar\lambda_+,D)$,
\begin{equation}
\def\arraystretch{1.3}
\begin{array}{rclcrcl}
 \delta A_1&=& -i\delta A_2=\xi_+\bar\lambda_++\bar\xi_+\lambda_+,&\quad&
 \delta\lambda_+ &=& \xi_+(iF_{12}+D), \\
 \delta D &=& (D_1+iD_2)(\xi_+\bar\lambda_+-\bar\xi_+\lambda_+),&\quad&
 \delta\bar\lambda_+ &=& \bar\xi_+(iF_{12}-D),
\end{array}
\def\arraystretch{1}
\end{equation}
chiral multiplet $(q,\psi_-)$ with its conjugate anti-chiral
multiplet $(\bar q,\bar\psi_-)$,
\begin{equation}
\def\arraystretch{1.3}
\begin{array}{rclcrcl}
 \delta q &=& 2\xi_+\psi_-, &\quad&
 \delta\psi_- &=& -\bar\xi_+(D_1+iD_2)q, \\
 \delta q &=& 2\bar\xi_+\bar\psi_-,&\quad&
 \delta\bar\psi_- &=& -\xi_+(D_1+iD_2)\bar q,
\end{array}
\def\arraystretch{1}
\end{equation}
and Fermi multiplet $(\psi_+,F;\Phi)$ with its conjugate
$(\bar\psi_+,\bar F;\bar\Phi)$,
\begin{equation}
\def\arraystretch{1.3}
\begin{array}{rcrcrcr}
 \delta\psi_+ &=& \xi_+F+\bar\xi_+\Phi, &\quad&
 \delta F &=& -2\bar\xi_+(D_1+iD_2)\psi_+ -2\bar\xi_+\Psi_-, \\
 \delta\bar\psi_+ &=& -\bar\xi_+\bar F+\xi_+\bar\Phi,&\quad&
 \delta\bar F &=&  2\xi_+(D_1+iD_2)\bar\psi_++2\xi_+\bar\Psi_-.
\end{array}
\def\arraystretch{1}
\end{equation}
Here $(\Phi,\Psi_-)$ is a chiral multiplet made of fields sitting in
other multiplets. The kinetic terms for these multiplets are
\begin{eqnarray}
 {\cal L}_\text{vec} &=&
 \text{Tr}\left(F_{12}^2+D^2-2\bar\lambda_+(D_1+iD_2)\lambda_+\right),
 \nonumber \\
 {\cal L}_\text{chi} &=&
 D_\mu\bar q D_\mu q+2\bar\psi_-(D_1-iD_2)\psi_-+i\bar q Dq
 -2i\bar\psi_-\bar\lambda_+q-2i\bar q\lambda_+\psi_-,
 \nonumber \\
 {\cal L}_\text{fer} &=&
 \bar\Phi\Phi+\bar FF-2\bar\psi_+(D_1+iD_2)\psi_+
 +2\bar\Psi_-\psi_+-2\bar\psi_+\Psi_-.
\label{Lkin}
\end{eqnarray}
All these are SUSY exact.
As usual, vector multiplet fields are regarded as matrices
and chiral (anti-chiral) fields are regarded as column (row) vectors.
Interactions can be introduced as an $F$-component of a gauge invariant
Fermi multiplet with vanishing $\Phi$-component. For example, consider
some chiral multiplets $(J^{(i)},\Xi^{(i)})$ and some Fermi multiplets
$(\psi^{(i)}_+,F^{(i)};\Phi^{(i)})$. Then $\sum_i\psi_+^{(i)}J^{(i)}$ is
the lowest component of a Fermi multiplet. From its $F$-component one
obtains
\begin{equation}
 {\cal L}_\text{int} ~=~ i\sum_i
 \Big(F^{(i)}J^{(i)}+\bar J^{(i)}\bar F^{(i)}
 -2\psi_+^{(i)}\Xi_-^{(i)}-2\bar\Xi_-^{(i)}\bar\psi_+^{(i)}\Big)\,,
\label{Lint}
\end{equation}
which is supersymmetric if $\sum_i\Phi^{(i)}J^{(i)}=0$. Another example
is the FI-theta term for abelian vector multiplets, which is indeed the
$F$-component of a Fermi multiplet starting from $\lambda_+$.

${\cal N}=(2,2)$ vector multiplet is obtained by combining an
${\cal N}=(0,2)$ vector multiplet $(A_\mu,\lambda_+,\bar\lambda_+,D)$
and an adjoint chiral multiplet $(Y,\lambda_-)$. Likewise, an ${\cal N}=(2,2)$
chiral multiplet is obtained by combining a chiral multiplet
$(q,\psi_-)$ and a Fermi multiplet $(\psi_+,F;\Phi=iYq)$ in the same
representation of the gauge group. The kinetic Lagrangian is given by a
sum of those in (\ref{Lkin}). To construct ${\cal L}_\text{int}$, one
chooses a gauge invariant function $W(q^{(i)})$ of chiral fields as
superpotential and set $J^{(i)}=\partial W/\partial q^{(i)}$ in
(\ref{Lint}).

${\cal N}=(4,4)$ vector multiplet consists of an ${\cal N}=(2,2)$ vector
multiplet $(A_\mu,Y,\lambda,\bar\lambda,D)$ and an adjoint chiral
multiplet $(\tilde Y,\tilde\lambda,F)$. A hypermultiplet is made of a
pair of ${\cal N}=(2,2)$ chiral multiplets, with lowest components
$q,\tilde q$, sitting in conjugate representations of gauge group.
The Lagrangian is uniquely determined from the gauge symmetry and its
representation. In particular we need to introduce a specific superpotential
$W=\tilde q\tilde Y q$ to have ${\cal N}=(4,4)$ supersymmetry.

\paragraph{Elliptic genus.}

Let us next introduce a powerful formula for elliptic genus of
${\cal N}=(0,2)$ supersymmetric gauge theories obtained in
\cite{Benini:2013nda,Benini:2013xpa}. Their derivation was based on
localization of path integral, and the final formula is expressed in
terms of the so-called Jeffrey-Kirwan residue.

For ${\cal N}=(0,2)$ theories on torus, all the SUSY invariants one can
use for Lagrangian are actually SUSY exact. For theories with standard
kinetic Lagrangians (\ref{Lkin}), the elliptic genus can be computed
using SUSY localization \cite{Benini:2013nda,Benini:2013xpa} and the
result basically depends only on the field content and symmetry.

Due to supersymmetry, the path integral localizes onto the moduli space
of BPS configurations, which in this case is the moduli space of flat gauge
fields. For rank-$r$ gauge group, the moduli space is real
$2r$-dimensional and is parametrized by $r$ complex coordinates
$w_1,\cdots,w_r$ with periodicity $w_i\sim w_i+1\sim w_r+\tau$. At first
glance, we seem to obtain an integral ${\rm d}^rw{\rm d}^r\bar w$ of
some one-loop determinant which is meromorphic in $w_i$. However, a
proper treatment of the zeromodes of the gaugino
$\lambda_+,\bar\lambda_+$ brings in a $\bar w_i$-dependence. It was
shown in \cite{Benini:2013nda,Benini:2013xpa} that, when this effect is
combined with a nice regularization of the divergence of the
determinant, the $2r$-dimensional integral can be transformed into an
integral of a certain $(r,0)$-form $Z_\text{1-loop}$ by a repeated use
of Stokes theorem. The elliptic genus thus becomes the sum of the
so-called Jeffrey-Kirwan residues of $Z_\text{1-loop}$ at all the
``poles'' $w_\ast$ in the moduli space where $r$ or more singular
hypersurfaces intersect.
\begin{equation}
 Z_{T^2}(\tau)=
 \frac1{|W|}\sum_{w_\ast}\underset{w=w_\ast}{\text{JK-Res}}(\eta)
 Z_\text{1-loop}(\tau,w).
\end{equation}
Here $|W|$ is the order of the Weyl group, and $\eta$ is a real
$r$-component vector we need to choose to define the residue operation.
The final result for $Z_{T^2}$ is independent of the choice of
$\eta$. The $(r,0)$-form $Z_\text{1-loop}$ is given by the product of
contributions from the vector, chiral and Fermi multiplets,
\begin{eqnarray}
 \Delta_\text{vec} &=&
 \Big(\frac{2\pi\eta(\tau)^2}{i}\Big)^r\prod_{\alpha:\text{roots}}
 \frac{i\theta_1(\tau|\alpha\cdot w)}{\eta(\tau)}\,{\rm d}^rw,
 \nonumber \\
 \Delta_\text{chi} &=&
 \prod_{\lambda:\text{weights}}
 \frac{i\eta(\tau)}{\theta_1(\tau|\lambda\cdot w)},
 \nonumber \\
 \Delta_\text{fer} &=&
 \prod_{\lambda:\text{weights}}
 \frac{i\theta_1(\tau|\lambda\cdot w)}{\eta(\tau)},
\label{Z1loops}
\end{eqnarray}
where $\theta_1$ is Jacobi theta function. It should be obvious how the
$Z_{T^2}$ will depend on additional parameters corresponding to external
flat gauge fields coupled to global symmetries.

\paragraph{Jeffrey-Kirwan residue formula.}

We next present some defining properties of the Jeffrey-Kirwan residue
operation which will be used in Section \ref{sec:EGABJM}.  A more detailed
definition is given in \cite{Benini:2013xpa}, see also \cite{SV,BV}.

Consider $n\,(\ge r)$ singular hyperplanes in $\mathbb R^r$ meeting at
the origin, defined by
\begin{equation}
 Q_i\cdot w=0.\quad(i=1,\cdots,n)
\end{equation}
We denote the ordered set of charges by $\Delta=\{Q_1,\cdots,Q_n\}$. In
the computation of elliptic genus, $Q_i$ are the weight vectors of the
$n$ matter chiral fields which acquire zeromodes at $w=0$. In the case
$n=r$, the Jeffrey-Kirwan residue is defined by the property
\begin{equation}
\def\arraystretch{1.2}
 \underset{w=0}{\text{JK-Res}}(\eta)
 \frac{{\rm d}^rw}
      {(Q_1\cdot w)\cdots (Q_r\cdot w)}
 = \left\{
   \begin{array}{ll}
    \displaystyle
    \frac1{|\text{det}(Q_1\cdots Q_r)|}~~ &
    \text{if}~\eta\in\text{Cone}(Q_1,\cdots,Q_r) \\
    0 & \text{otherwise}.
   \end{array}
   \right.
\def\arraystretch{1}~
\label{JKDef}
\end{equation}
Namely, it vanishes unless $\eta$ is expressed as a linear combination of
$\{Q_1,\cdots,Q_r\}$ with positive coefficients. Note that one should
keep track of the charge vectors $\{Q_i\}$ including their sign in the
residue computation.

When $n>r$, there is a number of ways to choose from $\Delta$ an ordered
set of $r$ linearly independent charge vectors
$b=\{Q_{i_1},\cdots,Q_{i_r}\}_{(i_1<\cdots<i_r)}$, which we call a basis
of $\Delta$. To each basis $b$ of $\Delta$ one can associate a
{\it basic fraction}
\begin{equation}
 \phi_b\equiv \frac1{\prod_{Q_i\in b}Q_i\cdot w}=
 \frac1{(Q_{i_1}\cdot w)\cdots (Q_{i_r}\cdot w)}.
\end{equation}
The basic fractions thus obtained may obey some
linear relations. For example, from the set of three charge vectors
$\Delta=\{(1,0),(0,1),(1,1)\}$ one gets three basic fractions
\begin{equation}
 \phi_{b_1}=\frac1{w_1w_2},~~
 \phi_{b_2}=\frac1{w_1(w_1+w_2)},~~
 \phi_{b_3}=\frac1{w_2(w_1+w_2)},
\end{equation}
obeying one relation $\phi_{b_1}=\phi_{b_2}+\phi_{b_3}$. In this example
the basic fractions form a 2-dimensional vector space, and any pair of
basic fractions can be used as a basis. We denote by $B$ a set of bases
of $\Delta$ such that $\{\phi_b\}_{b\in B}$ form a basis of basic
fractions. The Jeffrey-Kirwan residue of basic fractions is given by the
formula (\ref{JKDef}), whereas the fractions whose denominators do not
contain $r$ linearly independent factors, for example
\[
 \frac1{w_1},~ \frac1{(w_1+w_2)^2},~\cdots
\]
have trivial residues. More general meromorphic functions can all be
decomposed into derivatives of basis basic fractions and fractions with
trivial residue. For example,
\begin{equation}
 \frac1{w_1w_2(w_1+w_2)}= -\frac\partial{\partial w_1}\frac1{w_1w_2}
 +\Big(\frac\partial{\partial w_1}-\frac\partial{\partial w_2}\Big)
 \frac1{w_1(w_2+w_2)}.
\end{equation}
Therefore, the Jeffrey-Kirwan residue is uniquely determined by its
value on a basis of basic fractions.

Residue integrals can be regarded as linear functions on the space
of basic fractions. In the general case $n>r$ there are a number of
$r$-cycles to define residue integrals, and they are subject to some
linear relations. Here we quote from \cite{BV} a useful proposition. For
any $\Delta$, there is a choice of $B$ such that the set of iterated
residues $\text{Res}_{b}\;(b\in B)$
\begin{eqnarray}
 \text{Res}_b~\equiv~
 \underset{Q_{i_r}\!\cdot w=0}{\text{Res}}\cdots
 \underset{Q_{i_1}\!\cdot w=0}{\text{Res}}
 \qquad
\big(b=\{Q_{i_1},\cdots,Q_{i_r}\}_{(i_1<\cdots<i_r)}\in B\big)
\end{eqnarray}
form a dual basis to the basis of basic fractions $\{\phi_b\}_{b\in B}$,
namely $\text{Res}_b\phi_{b'}=\delta_{bb'}$. The iterated residue above
means one first takes the residue along the first hyperplane
$Q_{i_1}\cdot w=0$ keeping other $r-1$ variables fixed and generic, then
takes the residue along the second hyperplane $Q_{i_2}\cdot w=0$, and
goes on. Note that the order of the iterated residue is determined
according to the order of the charges in $\Delta$ we have chosen
(arbitrarily) at the beginning. With respect to this choice of $B$, the
Jeffrey-Kirwan residue is simply
\begin{equation}
 \text{Res}(\eta)= \sum_{b\in B;\;\eta\in\text{Cone}(b)}
 \nu(b)\cdot\text{Res}_b\,,
\end{equation}
where $\nu(b)=\pm1$ is the orientation of the basis $b$.

\subsection{IIA brane model}\label{sec:EG-IIA}

We first study the elliptic genus for the IIA brane model. As was
explained in Section \ref{sec:IIAbrane}, the theory can be obtained from
a 2D ${\cal N}=(4,4)$ supersymmetric gauge theory by freezing some of
the fields.

\TABLE{
\def\arraystretch{1.1}
\begin{tabular}{|c|c|c||cc|cc|}
\hline
& & &
\multicolumn{2}{|c|}{\small ${\cal N}\!\!=\!\!(0,2)$ chiral} &
\multicolumn{2}{|c|}{\small ${\cal N}\!\!=\!\!(0,2)$ Fermi} \\
\small${\cal N}\!\!=\!\!(4,4)$ & \small ${\cal N}\!\!=\!\!(2,2)$ &
\small rep &
\small(scalar) & \small(spinor) & \small(spinor) & \small(vector) \\ \hline
\multirow{4}{*}{vector} &
\multirow{2}{*}{vector} &
\multirow{2}{*}{adj} &
 $Y$ $(\ten\ten{+}{-})$&
 $\bar\lambda_-$ $(\ten{-}{+}\ten)$ &
 $\lambda_+$ $(\ten{+}\ten{-})$ &
 $A_\mu$  \\
 & &
 & $\bar Y$ $(\ten\ten{-}{+})$&
 $\lambda_-$ $(\ten{+}{-}\ten)$ &
 $\bar\lambda_+$ $(\ten{-}\ten{+})$ & \\ \cline{2-7}
 &\multirow{2}{*}{chiral} &
 \multirow{2}{*}{adj} &
 $\tilde Y$ $(\ten\ten{-}{-})$&
 $\tilde\lambda_-$ $(\ten{-}{-}\ten)$ &
 $\tilde\lambda_+$ $(\ten{-}\ten{-})$ & \\
 & & &
 $\bar{\tilde Y}$ $(\ten\ten{+}{+})$&
 $\bar{\tilde\lambda}_-$ $(\ten{+}{+}\ten)$ &
 $\bar{\tilde\lambda}_+$ $(\ten{+}\ten{+})$ & \\ \hline
\multirow{4}{*}{hyper} &
\multirow{2}{*}{chiral} &
 adj &
 $Z$ $({+}{+}\ten\ten)$&
 $\Psi_-$ $({+}\ten\ten{+})$ &
 $\Psi_+$ $({+}\ten{+}\ten)$ & \\
 & &
 adj &
 $\bar Z$ $({-}{-}\ten\ten)$&
 $\bar\Psi_-$ $({-}\ten\ten{-})$ &
 $\bar\Psi_+$ $({-}\ten{-}\ten)$ & \\ \cline{2-7}
 &
\multirow{2}{*}{chiral} &
 adj &
 $\tilde Z$ $({-}{+}\ten\ten)$&
 $\tilde\Psi_-$ $({-}\ten\ten{+})$ &
 $\tilde\Psi_+$ $({-}\ten{+}\ten)$ & \\
 & &
 adj &
 $\bar{\tilde Z}$ $({+}{-}\ten\ten)$&
 $\bar{\tilde{\Psi}}_-$ $({+}\ten\ten{-})$ &
 $\bar{\tilde{\Psi}}_+$ $({+}\ten{-}\ten)$ & \\ \hline
\multirow{4}{*}{hyper} &
\multirow{2}{*}{chiral} &
 $\Box$ &
 $q$ $(\ten{+}\ten\ten)$&
 $\psi_-$ $(\ten\ten\ten{+})$ &
 $\psi_+$ $(\ten\ten{+}\ten)$ & \\
 & &
 $\overline\Box$ &
 $\bar q$ $(\ten{-}\ten\ten)$&
 $\bar\psi_-$ $(\ten\ten\ten{-})$ &
 $\bar\psi_+$ $(\ten\ten{-}\ten)$ & \\ \cline{2-7}
 &
\multirow{2}{*}{chiral} &
 $\overline\Box$ &
 $\tilde q$ $(\ten{+}\ten\ten)$&
 $\tilde\psi_-$ $(\ten\ten\ten{+})$ &
 $\tilde\psi_+$ $(\ten\ten{+}\ten)$ & \\
 & &
 $\Box$ &
 $\bar{\tilde q}$ $(\ten{-}\ten\ten)$&
 $\bar{\tilde{\psi}}_-$ $(\ten\ten\ten{-})$ &
 $\bar{\tilde{\psi}}_+$ $(\ten\ten{-}\ten)$ & \\ \hline
\end{tabular}
\label{table:44theory}
\caption{The fields of an ${\cal N}=(4,4)$ SUSY theory with the $J^3$
 eigenvalues of the four $SU(2)$ global symmetries.
 The symbol $(\ten\ten{+}{-})$ means
 $(J^3_{(1)},J^3_{(2)},J^3_{(3)},J^3_{(4)})=(0,0,\frac12,-\frac12)$.
 In the ${\cal N}=(0,4)$ supersymmetric system of our interest, the
 fields in the multiplets including $Y,\tilde Y,\Psi_+,\tilde\Psi_+$ are
 frozen.}
}

The ${\cal N}=(4,4)$ system is a $U(N)$ gauge theory with one adjoint
and one fundamental hypermultiplets. The fields and their $J^3$ charges
under the global symmetry $SU(2)^4$ are listed in Table
\ref{table:44theory}. According to the charge assignments given there,
the parameter of ${\cal N}=(0,2)$ supersymmetry $\xi_+$ carries the charges
$(J^3_{(1)},J^3_{(2)},J^3_{(3)},J^3_{(4)})=(0,\frac12,0,-\frac12)$.
The ${\cal N}=(0,4)$ system of our interest is obtained by freezing the
multiplets including $Y,\tilde Y,\Psi_+$ and $\tilde\Psi_+$. If we turn on a
nonzero classical value $\langle Y\rangle =u\cdot{\bf 1}_{(N\times N)}$
in this process, the symmetry $SU(2)_3\times SU(2)_4$ is broken to a
diagonal $SU(2)$ which contains $J^3_{(3)}+J^3_{(4)}$.

In the standard approach reviewed above, the elliptic genus is given by
an integral with respect to the moduli of flat $U(N)$ gauge fields on
torus which we denote by $(w_1,\cdots, w_N)$. In
addition, we gauge the $U(1)^4$ symmetry generated by
$J^3_{(1)},\cdots,J^3_{(4)}$ by external flat gauge fields. The
corresponding fugacity parameters are denoted as follows,
\begin{equation}
 w_{\text{bg}(1)}=\epsilon_1-\epsilon_2,\quad
 w_{\text{bg}(2)}=\epsilon_1+\epsilon_2,\quad
 w_{\text{bg}(3)}=2m,\quad
 w_{\text{bg}(4)}=\epsilon_1+\epsilon_2.
\end{equation}
Note that we need $w_{\text{bg}(2)}=w_{\text{bg}(4)}$ in order to
preserve the ${\cal N}=(0,2)$ supersymmetry.
Using (\ref{Z1loops}) one finds that the meromorphic form
$Z_\text{1-loop}$ is given by
\begin{eqnarray}
 Z_{\text{1-loop}}
 &=& {\rm d}^Nw\times {\theta'_1}^N
 \prod_{i\ne j}\theta_1(w_i-w_j)
 \prod_{i,j}\frac{\theta_1(w_i-w_j+\epsilon_1+\epsilon_2)}
 {\theta_1(w_i-w_j+\epsilon_1)
  \theta_1(w_i-w_j+\epsilon_2)}
 \nonumber \\ &&\times
 \prod_i
 \frac{\theta_1(w_i-m)\theta_1(-w_i-m)}
      {\theta_1(w_i-\frac12(\epsilon_1+\epsilon_2))
       \theta_1(-w_i-\frac12(\epsilon_1+\epsilon_2))}\,.
\end{eqnarray}
Here and in the following, the first argument of $\theta_1$ function is
omitted since it is always $\tau$, and $\theta_1'=2\pi\eta(\tau)^3$ is
the $w$-derivative of $\theta_1(\tau|w)$ at $w=0$. If one wishes to turn
on a classical value of the adjoint scalar $Y$, an additional condition
$w_{\text{bg}(3)}=w_{\text{bg}(4)}$ or $2m=\epsilon_1+\epsilon_2$ is
needed due to symmetry breaking.

The evaluation of the Jeffrey-Kirwan residue seems quite intricate, so
we propose an alternative way to compute the elliptic genus based on
the idea of Higgs branch localization. This was proposed in the study of
$S^2$ partition function of 2D SUSY gauge theories in
\cite{Benini:2012ui,Doroud:2012xw}, and generalized to problems in
higher dimensions in
\cite{Chen:2013pha,Fujitsuka:2013fga,Benini:2013yva,Yoshida:2014qwa,Peelaers:2014ima}.
In our problem, this basically requires the BPS configurations to
satisfy also the F-term and D-term constraints arising from the equation
of motion of the auxiliary fields in ${\cal N}=(0,4)$ vector
multiplet. For the theory of our interest, the constraints are given by
\begin{eqnarray}
 q\bar q-\bar{\tilde q}\tilde q+[Z,\bar Z]+[\tilde Z,\bar{\tilde Z}]
 + \zeta\cdot{\bf 1}_{(N\times N)} &=& 0,
 \nonumber \\
 q\tilde q+[Z,\tilde Z] &=& 0,
\label{ADHM}
\end{eqnarray}
where we have turned on a FI deformation $\zeta$.
In fact, these are nothing but the ADHM equation for the moduli space of
$N$ $U(1)$-instantons. This is as expected since the D2-branes within a
D6-brane are known to behave like $U(1)$ instantons.

The constraints (\ref{ADHM}) and the $U(N)$ gauge equivalence determine
the moduli space of vacua of the theory. For nonzero $\zeta$, some
scalar fields must condense and break the gauge symmetry completely. In
the IR the theory flows to a non-linear sigma model on the moduli space
of $N$ $U(1)$-instantons, and what we are after is the elliptic genus for
that sigma model. In this sigma model the fugacity parameters
$\epsilon_1,\epsilon_2,m$ enter through the gauging of certain isometry
of the target space. In particular, there are a finite number of points on
the target space fixed under the isometry corresponding to
$\epsilon_1,\epsilon_2$. At each fixed point the sigma model is well
approximated by a free theory of chiral and Fermi multiplets coupled to
some external flat gauge fields. The elliptic genus for the sigma
model is obtained by summing over those fixed point contributions.

This consideration leads us to propose the following formula for the
elliptic genus
\begin{eqnarray}
 Z_\text{IIA}
 &=& \sum_{\{w_i\}}
 \prod_{i,j}
 \frac{\theta_1(w_i-w_j)\theta_1(w_i-w_j+\epsilon_1+\epsilon_2)}
      {\theta_1(w_i-w_j+\epsilon_1)\theta_1(w_i-w_j+\epsilon_2)}
 \nonumber \\ &&\hskip5mm\cdot
 \prod_i
 \frac{\theta_1(w_i-m)\theta_1(-w_i-m)}
      {\theta_1(w_i-\frac12(\epsilon_1+\epsilon_2))
       \theta_1(-w_i-\frac12(\epsilon_1+\epsilon_2))}\,.
\label{EG1HB}
\end{eqnarray}
Here the sum is over all the fixed points in Higgs branch, which are all
characterized by the value of flat $U(N)$ gauge field $\{w_i\}$. At each
of the fixed points, $\{w_i\}$ must be chosen appropriately so that some
scalars have zeromodes there and condense. The summand in the above
formula is the one-loop determinant without excluding the gaugino zeromodes.
It has therefore $N$ obvious zeroes $\theta_1(w_i-w_i)$ in the enumerator,
but at each fixed point we anticipate $N$ zeroes appear in the denominator
to cancel them. The formula (\ref{EG1HB}) makes sense and gives a finite
value once this cancellation is understood.

To determine the value of $\{w_i\}$ we notice that, in terms of the UV
gauge theory variables, the fixed points are described by the
solutions of (\ref{ADHM}) and the BPS condition
\begin{equation}
\def\arraystretch{1.3}
\begin{array}{rclcrcl}
 {\bf w}Z-Z{\bf w}+\epsilon_1Z &=& 0,&~~&
 {\bf w}q+\tfrac12(\epsilon_1+\epsilon_2)q &=& 0,\\
 {\bf w}\tilde Z-\tilde Z{\bf w}+\epsilon_2\tilde Z &=& 0, &&
-\tilde q{\bf w}+\tfrac12(\epsilon_1+\epsilon_2)\tilde q &=& 0,
\end{array}
\label{ADHMsad}
\end{equation}
where ${\bf w}\equiv\text{diag}(w_1,\cdots,w_N)$. It is an elementary
math problem to solve the combined system of equations (\ref{ADHM}) and
(\ref{ADHMsad}). See \cite{Kim:2011mv} for a detailed explanation.
For example, for $\zeta>0$ one can first show that
$\tilde q$ must be nonzero while $q=[Z,\tilde Z]=0$. Then one finds
there are $N$ linearly independent row vectors of the form
$\tilde qZ^m\tilde Z^n$ with the eigenvalue
${\bf w}=(m+\frac12)\epsilon_1+(n+\frac12)\epsilon_2$. The eigenvalue
spectrum is thus described by a Young diagram, so the fixed points are
labeled by a Young diagram of $N$ boxes. The precise value of the
scalar fields at the fixed points is not important. But a careful look
at the solution shows that, among $2N^2+2N$ components of scalar fields
$(Z,\tilde Z,q,\tilde q)$, there are precisely $N$ components acquiring
nonzero values at every fixed point.

To find out the free theory that approximates the sigma model at each
fixed point and compute its elliptic genus, we simply substitute the
eigenvalues of ${\bf w}$ into the summand of (\ref{EG1HB}). Since $\{w_i\}$
have been chosen so that $N$ scalars can condense, the denominator of
the summand in (\ref{EG1HB}) gives rise to $N$ zeroes that precisely cancel the
$N$ zeroes in the enumerator, as expected. This is only a part of the
manifestation of super Higgs mechanism; in fact, many similar cancellations
occur for theta functions with nonzero arguments as well. The
contribution from each fixed point can be evaluated using the formula
\cite{Nakajima:2003pg}
\begin{eqnarray}
  \sum_{(k,l)\in{\bf Y}}t_1^kt_2^l
 +\sum_{(k',l')\in{\bf Y}'}t_1^{1-k'}t_2^{1-l'}
 -\sum_{(k,l)\in{\bf Y},\;(k',l')\in{\bf Y}'}t_1^{k-k'}t_2^{l-l'}(1-t_1)(1-t_2)
 \nonumber \\ ~=~
 \sum_{(k,l)\in{\bf Y}}
  t_1^{k-\tilde\lambda_l({\bf Y}')}
  t_2^{1+\lambda_k({\bf Y})-l}
+\sum_{(k',l')\in{\bf Y}'}
  t_1^{1+\tilde\lambda_{l'}({\bf Y})-k'}
  t_2^{l'-\lambda_{k'}({\bf Y}')}\,,
\end{eqnarray}
with the sum of monomials translated into the product of theta functions.
Here $(k,l)\in{\bf Y}$ is a pair of positive integers labeling a
box in the diagram ${\bf Y}$, and
$\lambda_k({\bf Y}),\tilde\lambda_l({\bf Y})$ are the lengths of its $k$-th
column and $l$-th row. Thus the contribution to the elliptic genus from
the fixed point labeled by a Young diagram ${\bf Y}$ is
\begin{equation}
 Z_{\bf Y}= \prod_{(i,j)\in{\bf Y}}
 \frac
{\theta_1\left(-m+(i-\frac12)\epsilon_1+(j-\frac12)\epsilon_2\right)\cdot
 \theta_1\left(-m-(i-\frac12)\epsilon_1-(j-\frac12)\epsilon_2\right)}
{\theta_1\big((i-\tilde\lambda_j)\epsilon_1+(\lambda_i-j+1)\epsilon_2\big)
 \cdot
 \theta_1\big((\tilde\lambda_j-i+1)\epsilon_1+(j-\lambda_i)\epsilon_2\big)}.
\end{equation}
This formula can be interpreted as the elliptic genus of a
free theory of $2N$ chiral multiplets and $2N$ Fermi multiplets coupled
to some background flat gauge field, and the number $2N$ agrees with the
complex dimension of the Higgs branch moduli space. The elliptic genus
of the IIA brane model is finally given by
\begin{equation}
 Z_\text{IIA} ~=~ \sum_{\bf Y}Z_{\bf Y}\,.
\label{EG1}
\end{equation}
This agrees with the result obtained from topological vertex formalism
\cite{Haghighat:2013gba}.

So far we have been considering the case with
$\langle Y\rangle=\langle\tilde Y\rangle=0$.
If $2m=\epsilon_1+\epsilon_2$, one can turn on $\langle Y\rangle$ and
make the fundamental fields massive. The elliptic genus for such theory
should be independent of the mass of fundamental matters. In the limit
$\langle Y\rangle\to\infty$ the fundamental fields are frozen and we are
left with ${\cal N}=(4,4)$ $U(N)$ super Yang-Mills theory. Note that, as
explained in Section \ref{sec:IIAbrane}, the ${\cal N}=(4,4)$ enhanced
SUSY here is different from the one used to classify the fields in Table
\ref{table:44theory}, and that the fields $A_\mu,Z,\tilde Z$ form the
${\cal N}=(4,4)$ vector multiplet together with the fermions. Since all
the remaining degrees of freedom are in the adjoint, the $U(1)$ part of
the vector multiplet becomes free, and in particular has fermion
zeromodes. The contribution to the elliptic genus from the $U(1)$ part
is identified with $Z_\text{IIA}$ for the case $N=1$. Before setting
$2m=\epsilon_1+\epsilon_2$ it is given by \cite{Haghighat:2013gba}
\begin{equation}
 Z_{\text{IIA}(N=1)}=
 \frac{\theta_1(m+\frac12\epsilon_1+\frac12\epsilon_2)
       \theta_1(m-\frac12\epsilon_1-\frac12\epsilon_2)}
      {\theta_1(\epsilon_1)\theta_1(\epsilon_2)}.
\label{ZU1-IIA}
\end{equation}
Factoring this out from $Z_\text{IIA}$ and setting
$m=\frac12(\epsilon_1+\epsilon_2)$, we should obtain the elliptic genus
for ${\cal N}=(4,4)$ $SU(N)$ super Yang-Mills theory.
\begin{equation}
 \frac{Z_\text{IIA}}{Z_{\text{IIA}(N=1)}}\bigg|_{m=\frac12(\epsilon_1+\epsilon_2)}
 ~=~ Z_{SU(N)}.
\end{equation}

\subsection{ABJM slab}\label{sec:EGABJM}

Next we study the elliptic genus for the theory (\ref{44TYM}) obtained
from the dimensional reduction of the ABJM model. Although the
importance of KK modes was emphasized in Section \ref{sec:qYM}, those
massive modes can be safely neglected for the computation of the
elliptic genus. Also, we set here $k=1$ to avoid the complication with
the $\mathbb Z_k$ orbifolding.

The matter fields, namely $Z$'s and $\Psi$'s in the Lagrangian
(\ref{44TYM}) are organized into two chiral and two Fermi multiplets,
all sitting in the adjoint representation. The gauge field is promoted
to a $U(N)$ vector multiplet $(A_\mu,\lambda_+,\bar\lambda_+,D)$,
whereas the scalar $\sigma$ is promoted to an adjoint chiral multiplet
$(Y,\bar\lambda_-)$ with the lowest component $Y=\rho+i\sigma$. The
off-shell ${\cal N}=(0,2)$ supersymmetric version of the topological
Yang-Mills Lagrangian is
\begin{equation}
 {\cal L}_\text{top} = \frac{ik}{2\pi}\text{Tr}
 \Big(\sigma F_{12}-\rho D +\lambda_+\bar\lambda_-+\lambda_-\bar\lambda_+\Big).
\end{equation}
The fields and their quantum numbers under the global symmetry $SU(2)^4$
are summarized in the Table \ref{table:2}.

\TABLE[t]{
\def\arraystretch{1.1}
\begin{tabular}{|cc|cc|}
\hline
\multicolumn{2}{|c|}{\small ${\cal N}\!\!=\!\!(0,2)$ chiral} &
\multicolumn{2}{|c|}{\small ${\cal N}\!\!=\!\!(0,2)$ Fermi} \\
\small(scalar) & \small(spinor) & \small(spinor) & \small(vector) \\ \hline
 $Y$ $(\ten\ten\ten\ten)$&
 $\bar\lambda_-$ $(\ten{-}\ten{+})$ &
 $\lambda_+$ $(\ten{+}\ten{-})$ &
 $A_\mu$  \\
 $\bar Y$ $(\ten\ten\ten\ten)$&
 $\lambda_-$ $(\ten{+}\ten{-})$ &
 $\bar\lambda_+$ $(\ten{-}\ten{+})$ & \\ \hline
 $Z$ $({+}{+}\ten\ten)$&
 $\Psi_-$ $({+}\ten\ten{+})$ &
 $\Psi_+$ $(\ten{+}{+}\ten)$ & \\
 $\bar Z$ $({-}{-}\ten\ten)$&
 $\bar\Psi_-$ $({-}\ten\ten{-})$ &
 $\bar\Psi_+$ $(\ten{-}{-}\ten)$ & \\ \hline
 $\tilde Z$ $({-}{+}\ten\ten)$&
 $\tilde\Psi_-$ $({-}\ten\ten{+})$ &
 $\tilde\Psi_+$ $(\ten{-}{+}\ten)$ & \\
 $\bar{\tilde Z}$ $({+}{-}\ten\ten)$&
 $\bar{\tilde{\Psi}}_-$ $({+}\ten\ten{-})$ &
 $\bar{\tilde{\Psi}}_+$ $(\ten{+}{-}\ten)$ & \\ \hline
\end{tabular}
\label{table:2}
\caption{The fields of the dimensionally reduced ABJM model at $u=0$
 with the $J^3$ eigenvalues of the four $SU(2)$ global symmetries. All
 fields are in the adjoint of the gauge group $U(N)$.}
}

Since the gaugino $\lambda_+$ in this model is an auxiliary field, we
can try the SUSY path integral using the original Lagrangian
${\cal L}_\text{top}$, not the usual kinetic Lagrangians (\ref{Lkin}),
for the path integral weight for the multiplets containing $A_\mu$ and
$\sigma$. The path integral of these multiplets is then trivial, and we
are left with an integral over the moduli of flat $U(N)$ gauge fields
and the 1-loop determinant arising only from the fields
$Z,\tilde Z,\Psi_\pm$ and $\tilde\Psi_\pm$. The elliptic genus would
then be given by
\begin{equation}
 Z_{T^2}\sim \int\prod_{i=1}^N
 \frac{{\rm d}w_i{\rm d}\bar w_i}{\text{Im}\tau}\cdot
 \prod_{i,j}\frac
 {\theta_1(w_i-w_j+m+\frac12(\epsilon_1+\epsilon_2))
  \theta_1(w_i-w_j+m-\frac12(\epsilon_1+\epsilon_2))}
 {\theta_1(w_i-w_j+\epsilon_1)\theta_1(w_i-w_j+\epsilon_2)}\,.
\label{EGslab}
\end{equation}
However, there is a priori no natural way to rewrite it further into some
integral of an $(N,0)$-form. Similar formulae for the elliptic genus
have been proposed for theories with St\"uckelberg fields in
\cite{Ashok:2013pya,Murthy:2013mya}, though in those cases the elliptic
genus becomes a non holomorphic function of $\tau$.

The integral with respect to $w_i,\bar w_i$ is finite and does not
require careful regularization. For $N=1$ the integrand does not depend
on $w_1,\bar w_1$ at all, so the above $Z_{T^2}$ agrees with the
result of IIA brane model (\ref{ZU1-IIA}). For higher $N$, the integrand
contains the $N$-power of the $U(1)$ part (\ref{ZU1-IIA}), so it vanishes
faster than the elliptic genus of the IIA brane model as
$m\to\frac12(\epsilon_1+\epsilon_2)$. This is due to the Fermi
multiplet $\tilde\Psi_+$ which acquire $N$ zeromodes in this limit.
Thus the elliptic genera of the IIA brane model and that of ABJM slab
do not agree.

When $2m=\epsilon_1+\epsilon_2$, then the Fermi multiplet $\tilde\Psi_+$
couples to no external gauge fields, so one can consider turning off
${\cal L}_\text{top}$ in the original Lagrangian and instead add
standard kinetic term ${\cal L}_\text{vec}$ for the vector multiplet, and
also a mass term which is bilinear in the multiplets $Y$ and
$\tilde\Psi_+$. After neglecting the fields which become massive, the
remaining field content is the same as that of ${\cal N}=(4,4)$ $U(N)$
super Yang-Mills theory. This deformation should correspond to turning
on $u_A$ and integrating out the scalar $\sigma$ in Section
\ref{sec:ABJMslab}. Although the deformation is SUSY exact, it involves
the multiplets with zeromodes and therefore changes the asymptotic
behavior of path integral weight. Under such deformation the elliptic
genus may well change. Indeed, while (\ref{EGslab}) gives rise to $N$
powers of zeroes at $m=\frac12(\epsilon_1+\epsilon_2)$, the elliptic
genus for $U(N)$ super Yang-Mills theory has only one zero.

\paragraph{Young diagram sum from Jeffrey-Kirwan residue.}

After setting $m=\frac12(\epsilon_1+\epsilon_2)$ and removing the $U(1)$
part (which is vanishing due to a fermion zeromode), the elliptic genus
of the remaining $SU(N)$ super Yang-Mills theory is given by the
Jeffrey-Kirwan residue of the following meromorphic form,
\begin{equation}
 Z_\text{1-loop} =
 \prod_{i=1}^{N-1}{\rm d}\hat w_i\cdot\frac1{N!}
 \Big(\frac{\theta_1'\theta_1(\epsilon_1+\epsilon_2)}
           {\theta_1(\epsilon_1)\theta_1(\epsilon_2)}\Big)^{N-1}
 \prod_{i\ne j}\frac
 {\theta_1(w_i-w_j)\theta_1(w_i-w_j-\epsilon_1-\epsilon_2)}
 {\theta_1(w_i-w_j-\epsilon_1)\theta_1(w_i-w_j-\epsilon_2)}\,.
\label{Z1L-ABJM}
\end{equation}
Here the coordinates $w_i$ are assumed to satisfy $\sum_iw_i=0$. More
explicitly, they are expressed in terms of the moduli of flat $SU(N)$
gauge fields $\hat w_i$ as follows,
\begin{equation}
 \left(w_1,\cdots,w_N\right)=
 \left(-\hat w_1,\;\hat w_1-\hat w_2,\;\hat w_2-\hat w_3,\;\cdots,\;
 \hat w_{N-2}-\hat w_{N-1},\;\hat w_{N-1}\right)
\label{UNvsSUN}
\end{equation}
The coordinates $\hat w_i$ obey the periodicity
$\hat w_i\sim \hat w_i+1\sim\hat w_i+\tau$.

In the previous subsection we have shown that the elliptic genus
of ${\cal N}=(4,4)$ super Yang-Mills theory is expressed as a sum over
contributions labeled by Young diagrams, namely (\ref{EG1}) with
$m=\frac12(\epsilon_1+\epsilon_2)$ substituted and the $U(1)$ part
removed. The derivation there was based on the Higgs branch localization
in a system with fundamental matters. Here we wish to re-derive the same
result as the Jeffrey-Kirwan residue of the meromorphic form
(\ref{Z1L-ABJM}).

The singular hyperplanes of $Z_\text{1-loop}$ (\ref{Z1L-ABJM}) are given
by $w_i-w_j=\epsilon_1$ or $w_i-w_j=\epsilon_2$. It is convenient to
describe their intersections graphically by arrangements of $N$
particles on a 2D square lattice. For example, some poles in the case
$N=4$ and the corresponding graphs are
\setlength\unitlength{15pt}
\begin{eqnarray}
 \{w_2-w_1=\epsilon_1,~w_3-w_2=\epsilon_1,~w_4-w_3=\epsilon_1\}
 &\Longleftrightarrow&
\begin{picture}(4,0.5)(-0.5,-0.1)
\put(0,0){\num{1}} \put(0,0)\dt \put(0,0)\hl
\put(1,0){\num{2}} \put(1,0)\dt \put(1,0)\hl
\put(2,0){\num{3}} \put(2,0)\dt \put(2,0)\hl
\put(3,0){\num{4}} \put(3,0)\dt
\end{picture}
 \nonumber \\
 \{w_2-w_1=\epsilon_1,~w_3-w_1=\epsilon_2,~w_4-w_2=\epsilon_1\}
 &\Longleftrightarrow&
\begin{picture}(3,1.5)(-0.5,0.3)
\put(0,0){\num{1}} \put(0,0)\dt \put(0,0)\hl \put(0,0)\vl
\put(1,0){\num{2}} \put(1,0)\dt \put(1,0)\hl
\put(0,1){\num{3}} \put(0,1)\dt
\put(2,0){\num{4}} \put(2,0)\dt
\end{picture}
\label{PA-YD}
\end{eqnarray}
Note that the singular hyperplanes correspond to the links connecting
particles occupying the neighboring sites. Some other poles, for example
\begin{eqnarray}
 \{w_2-w_1=\epsilon_1,~w_3-w_1=\epsilon_1,~
   w_4-w_2=\epsilon_1,~w_4-w_3=\epsilon_1\}
 &\Longleftrightarrow&
\begin{picture}(3,0.5)(-0.5,-0.1)
\put(0,0){\num{1}}     \put(0,0)\dt \put(0,0)\hl
\put(1,0){\num{~~2,3}} \put(1,0)\dt \put(1,0)\hl
\put(2,0){\num{4}}     \put(2,0)\dt
\end{picture}
 \nonumber \\
 \{w_2-w_1=\epsilon_1,~w_3-w_2=\epsilon_1,~w_4-w_3=\epsilon_2\}
 &\Longleftrightarrow&
\begin{picture}(3,1.5)(-0.5,0.3)
\put(0,0){\num{1}} \put(0,0)\dt \put(0,0)\hl
\put(1,0){\num{2}} \put(1,0)\dt \put(1,0)\hl \put(1,0)\dl
\put(2,0){\num{3}} \put(2,0)\dt \put(2,0)\vl
\put(2,1){\num{4}} \put(2,1)\dt
\end{picture}
\label{PA-1stq}
\end{eqnarray}
\vskip1.5mm\noindent
will have vanishing residues due to the factors in the enumerator of
$Z_\text{1-loop}$. In the first example where $w_2$ and $w_3$ occupy the
same lattice site, the residue vanishes due to the factor
$\theta_1(w_2-w_3)^2$ although 4 hyperplanes are intersecting
there. In the second example the residue vanishes due to the enumerator factor
$\theta_1(w_4-w_2-\epsilon_1-\epsilon_2)$ which is indicated by the
slash (thin oblique line) in the graph.

\FIGURE{
\begin{picture}(4,3.5)(-.5,-.5)
\put(0,0){\num{8}} \put(0,0)\dt \put(0,0)\hl \put(0,0)\vl \put(0,0)\dl
\put(1,0){\num{9}} \put(1,0)\dt \put(1,0)\hl \put(1,0)\vl \put(1,0)\dl
\put(2,0){\num{1}} \put(2,0)\dt \put(2,0)\hl \put(2,0)\vl
\put(3,0){\num{3}} \put(3,0)\dt
\put(0,1){\num{4}} \put(0,1)\dt \put(0,1)\hl \put(0,1)\vl \put(0,1)\dl
\put(1,1){\num{2}} \put(1,1)\dt \put(1,1)\hl \put(1,1)\vl
\put(2,1){\num{5}} \put(2,1)\dt
\put(0,2){\num{7}} \put(0,2)\dt \put(0,2)\hl
\put(1,2){\num{6}} \put(1,2)\dt
\end{picture}
}
Let us first focus on the poles corresponding to Young tableaux (= Young
diagrams with numbering of boxes). The figure on the right shows an
example of a pole for the case $N=9$, where 11 hypersurfaces
(corresponding to 5 vertical and 6 horizontal links) intersect, and 3
zeroes arising from the enumerator. For a pole corresponding
to a general Young tableau ${\bf T}$, we consider the simplified
meromorphic form,
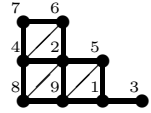
\begin{equation}
 \hat Z_\text{1-loop}({\bf T}) = {\rm d}^{N-1}\hat w\cdot
\frac{\displaystyle
 \prod_{\scriptscriptstyle\nearrow}(w_j-w_i-\epsilon_1-\epsilon_2)}
{\displaystyle
 \prod_{\scriptscriptstyle\rightarrow}(w_j-w_i-\epsilon_1)\cdot
 \prod_{\scriptscriptstyle\uparrow}(w_j-w_i-\epsilon_2)},
\label{Zhat}
\end{equation}
which extracts the singular and vanishing factors from $Z_\text{1-loop}$
in (\ref{Z1L-ABJM}).
The three products are over the slashes, horizontal links and
vertical links, respectively. Note that $\hat Z_\text{1-loop}$ has
degree $N-1$ for any tableau ${\bf T}$ of $N$ boxes, namely
$\hat Z_\text{1-loop}$ has always $N-1$ more factors in the denominator
than in the enumerator.

We use the following identities
\setlength\unitlength{12pt}
\begin{equation}
\begin{picture}(1.6,.6)(-0.3,0.3)
\put(0,0)\dt \put(0,0)\hl \put(0,0)\dl
\put(1,0)\dt \put(1,0)\vl
\put(1,1)\dt
\end{picture}
=
\begin{picture}(1.6,.6)(-0.3,0.3)
\put(0,0)\dt
\put(1,0)\dt \put(1,0)\vl
\put(1,1)\dt
\end{picture}
+
\begin{picture}(1.6,.6)(-0.3,0.3)
\put(0,0)\dt \put(0,0)\hl
\put(1,0)\dt
\put(1,1)\dt
\end{picture}
~~,\qquad
\begin{picture}(1.6,.6)(-0.3,0.3)
\put(0,0)\dt \put(0,0)\vl \put(0,0)\dl
\put(0,1)\dt \put(0,1)\hl
\put(1,1)\dt
\end{picture}
=
\begin{picture}(1.6,.6)(-0.3,0.3)
\put(0,0)\dt
\put(0,1)\dt \put(0,1)\hl
\put(1,1)\dt
\end{picture}
+
\begin{picture}(1.6,.6)(-0.3,0.3)
\put(0,0)\dt \put(0,0)\vl
\put(0,1)\dt
\put(1,1)\dt
\end{picture}
\label{wid}
\end{equation}
to rewrite $\hat Z_\text{1-loop}({\bf T})$ into a sum of basic
fractions. Here is an $N=6$ example (the numbering of particles is suppressed).
\begin{eqnarray}
\begin{picture}(2.6,1.5)(-0.3,0.3)
\put(0,0)\dt \put(0,0)\hl \put(0,0)\vl \put(0,0)\dl
\put(1,0)\dt \put(1,0)\hl \put(1,0)\vl \put(1,0)\dl
\put(2,0)\dt \put(2,0)\vl
\put(0,1)\dt \put(0,1)\hl
\put(1,1)\dt \put(1,1)\hl
\put(2,1)\dt
\end{picture}
 &=&
\begin{picture}(2.6,1.5)(-0.3,0.3)
\put(0,0)\dt \put(0,0)\vl
\put(1,0)\dt \put(1,0)\vl
\put(2,0)\dt \put(2,0)\vl
\put(0,1)\dt \put(0,1)\hl
\put(1,1)\dt \put(1,1)\hl
\put(2,1)\dt
\end{picture}
+
\begin{picture}(2.6,1.5)(-0.3,0.3)
\put(0,0)\dt \put(0,0)\vl
\put(1,0)\dt \put(1,0)\hl \put(1,0)\vl
\put(2,0)\dt
\put(0,1)\dt \put(0,1)\hl
\put(1,1)\dt \put(1,1)\hl
\put(2,1)\dt
\end{picture}
+
\begin{picture}(2.6,1.5)(-0.3,0.3)
\put(0,0)\dt \put(0,0)\hl \put(0,0)\vl
\put(1,0)\dt
\put(2,0)\dt \put(2,0)\vl
\put(0,1)\dt \put(0,1)\hl
\put(1,1)\dt \put(1,1)\hl
\put(2,1)\dt
\end{picture}
+
\begin{picture}(2.6,1.5)(-0.3,0.3)
\put(0,0)\dt \put(0,0)\hl \put(0,0)\vl
\put(1,0)\dt \put(1,0)\hl
\put(2,0)\dt
\put(0,1)\dt \put(0,1)\hl
\put(1,1)\dt \put(1,1)\hl
\put(2,1)\dt
\end{picture}
 \nonumber \\
 &=&
\begin{picture}(2.6,1.5)(-0.3,0.3)
\put(0,0)\dt \put(0,0)\hl
\put(1,0)\dt \put(1,0)\vl
\put(2,0)\dt \put(2,0)\vl
\put(0,1)\dt \put(0,1)\hl
\put(1,1)\dt \put(1,1)\hl
\put(2,1)\dt
\end{picture}
+
\begin{picture}(2.6,1.5)(-0.3,0.3)
\put(0,0)\dt \put(0,0)\hl
\put(1,0)\dt \put(1,0)\hl \put(1,0)\vl
\put(2,0)\dt
\put(0,1)\dt \put(0,1)\hl
\put(1,1)\dt \put(1,1)\hl
\put(2,1)\dt
\end{picture}
+
\begin{picture}(2.6,1.5)(-0.3,0.3)
\put(0,0)\dt \put(0,0)\hl \put(0,0)\vl
\put(1,0)\dt \put(1,0)\vl
\put(2,0)\dt \put(2,0)\vl
\put(0,1)\dt
\put(1,1)\dt \put(1,1)\hl
\put(2,1)\dt
\end{picture}
+
\begin{picture}(2.6,1.5)(-0.3,0.3)
\put(0,0)\dt \put(0,0)\hl \put(0,0)\vl
\put(1,0)\dt \put(1,0)\hl \put(1,0)\vl
\put(2,0)\dt
\put(0,1)\dt
\put(1,1)\dt \put(1,1)\hl
\put(2,1)\dt
\end{picture}
\label{Z-BF}
\end{eqnarray}
Each basic fraction is described by a {\it tree graph} of $N-1$ links
connecting $N$ particles. Note that the
decomposition of $\hat Z_\text{1-loop}({\bf T})$ into basic fractions is
in general not unique. Once the numbering of $N$ particles is restored,
one can associate to each basic fraction a set $b$ of $N-1$ charge
vectors ($SU(N)$ roots). For example we have,
\setlength\unitlength{15pt}
\begin{equation}
\begin{picture}(2.6,1.5)(-0.3,0.3)
\put(0,0){\num{1}} \put(0,0)\dt \put(0,0)\vl
\put(1,0){\num{2}} \put(1,0)\dt \put(1,0)\hl \put(1,0)\vl
\put(2,0){\num{3}} \put(2,0)\dt
\put(0,1){\num{4}} \put(0,1)\dt \put(0,1)\hl
\put(1,1){\num{5}} \put(1,1)\dt \put(1,1)\hl
\put(2,1){\num{6}} \put(2,1)\dt
\end{picture}
~\Longrightarrow~
b=:\{\alpha_{14},\alpha_{45},\alpha_{25},\alpha_{23},\alpha_{56}\}\,.
\label{treegraph}
\end{equation}
\vskip1.5mm\noindent
Here we identified the $SU(N)$ root $\alpha_{ij}$ with the $N$-component
vector whose $i$-th component is $-1$, $j$-th component $+1$ and others
zero, so that $\alpha_{ij}\cdot w=w_j-w_i$.

To evaluate the Jeffrey-Kirwan residue, we choose $\eta$ to be a
generic $N$-component vector satisfying
\begin{equation}
 \eta_1=-(\eta_2+\cdots+\eta_N),\quad
 \eta_2>0,\cdots,\eta_N>0.
\label{eta}
\end{equation}
For each tree graph such as (\ref{treegraph}), we need to determine
whether $\eta$ is inside the cone generated by the charge vectors in $b$.
To do this graphically, we think of a current flowing
along the tree graph. The lattice site occupied by the $i$-th particle
($i\ne 1$) is a source of strength $\eta_i$, and the site occupied by
the $1$-st particle is a sink of strength $(\eta_2+\cdots+\eta_N)$. We
then compute the flow along each link. For example, for the graph given
in (\ref{treegraph}) we find
\setlength\unitlength{15pt}
\begin{eqnarray}
\begin{picture}(5.5,2.2)(-0.3,0)
\put(0,0)\dt \put(0,0)\vl
\put(1,0)\dt \put(1,0)\hl \put(1,0)\vl
\put(2,0)\dt
\put(0,1)\dt \put(0,1)\hl
\put(1,1)\dt \put(1,1)\hl
\put(2,1)\dt
\put(2.7,1){\vector(-1,0){.5}}
\put(2.8,.95){\scriptsize $\eta_6$}
\put(2.7,0){\vector(-1,0){.5}}
\put(2.8,-0.05){\scriptsize $\eta_3$}
\put(0,1.7){\vector(0,-1){0.5}}
\put(-.2,1.9){\scriptsize $\eta_4$}
\put(1,1.7){\vector(0,-1){0.5}}
\put(.8,1.9){\scriptsize $\eta_5$}
\put(1,-.7){\vector(0,1){0.5}}
\put(.8,-1){\scriptsize $\eta_2$}
\put(0,-.2){\vector(0,-1){1.2}}
\put(-.2,-1.7){\scriptsize $\eta_2+\eta_3+\eta_4+\eta_5+\eta_6$}
\end{picture}
\Longrightarrow~~\eta &=&
 (\eta_2+\eta_3+\eta_4+\eta_5+\eta_6)\alpha_{14}
+(\eta_2+\eta_3+\eta_5+\eta_6)\alpha_{45}
 \nonumber \\ &&
-(\eta_2+\eta_3)\alpha_{25}
+\eta_3\alpha_{23}
+\eta_6\alpha_{56}.
\label{etadcmp}
\end{eqnarray}
\vskip1.5mm\noindent
This shows $\eta$ is outside of the cone generated by $b$ of
(\ref{treegraph}) because the coefficient of $\alpha_{25}$ is negative.

\FIGURE{
\begin{picture}(5.3,2.2)(-.5,0)
\put(0,0)\dt \put(0,0)\vl
\put(1,0)\dt \put(1,0)\hl \put(1,0)\vl
\put(2,0)\dt
\put(0,1)\dt \put(0,1)\hl
\put(1,1)\dt \put(1,1)\hl
\put(2,1)\dt
\put(-.5,-.5){\vector(1,0){4.7}}
\put(0.5,-.7){\line(0,1){.4}}
\put(1.5,-.7){\line(0,1){.4}}
\put(2.5,-.7){\line(0,1){.4}}
\put(3.5,-.7){\line(0,1){.4}}
\put(0.5,-.5){\line(-1,1){1}}
\put(1.5,-.5){\line(-1,1){2}}
\put(2.5,-.5){\line(-1,1){2}}
\put(3.5,-.5){\line(-1,1){2}}
\put(0.5,-1.1){\tiny$0$}
\put(1.5,-1.1){\tiny$1$}
\put(2.5,-1.1){\tiny$2$}
\put(3.5,-1.1){\tiny$3$}
\put(4.0,-.2){\tiny$h$}
\end{picture}
}

In order to make more general statements, it is useful to introduce the
height function $h$ on tree graphs. As shown in the figure on the
right, we assign height $0$ to the unique bottom-left corner site of the
graph. The height increases (decreases) by one as we move along the tree
graph one step up or right (down or left). Looking back
at the computation (\ref{etadcmp}), we notice that the minus sign in a
coefficient corresponds to a backward flow of current along the link
between $2$ and $5$. For $\eta$ to be inside the cone, the tree graph
and the arrangement of $N$ particles on it should have been such that the
current be flowing everywhere in accordance with the height gradient.

For a basic fraction to have nonzero Jeffrey-Kirwan residue, the
corresponding tree graph has to be {\it monotonic}, which means the
following. The 1-st particle is at the lowest site, and the height $h$
increases as one goes away from the 1-st particle along any path on the
tree graph until one reaches an end. For the $\hat Z_\text{1-loop}({\bf
T})$ to have nonzero Jeffrey-Kirwan residue,
its decomposition into basic fractions must contain a term corresponding
to monotonic tree graph, therefore the number $1$ must be assigned to the
bottom-left corner of ${\bf T}$. In the sample decomposition of
$\hat Z_\text{1-loop}$ into basic fractions (\ref{Z-BF}), we
notice that in each line there is only one monotonic tree graph, namely
the last term in each line. One can actually show that the decomposition
always gives rise to only one monotonic tree graph irrespective of
${\bf T}$ or the ways of decomposition, as follows. The only monotonic
tree graph can be obtained by choosing, when applying the identities
(\ref{wid}), always the second term on the RHS. An important characteristic of
monotonic tree graph is that, if it has $N_h$ points at height $h$, it
has exactly $N_h$ links between points of heights $h-1$ and $h$. This
property is violated once one makes the other choice when applying
(\ref{wid}).

The Jeffrey-Kirwan residue of a basic fraction corresponding to a tree
diagram is, if nonzero, always given by the inverse of the determinant
of $SU(N)$ Cartan matrix.
\begin{equation}
 \text{JK-Res}(\eta)\left[
 \frac{{\rm d}^{N-1}\hat w}{\prod_{\alpha\in b}\alpha\cdot w}\right]
 =\left\{\begin{array}{ll}
   N^{-1} ~~& \text{if}~\eta\in\text{Cone}(b) \\
   0   & \text{otherwise}. \end{array}
   \right.
\end{equation}
Thus we have, for any Young tableau ${\bf T}$,
\begin{equation}
 \text{JK-Res}(\eta)\hat Z_\text{1-loop}({\bf T})
 = \left\{\begin{array}{ll}
   N^{-1}~~ & \text{if 1 is in the bottom-left corner of {\bf T}}\\
   0 & \text{otherwise}.
	  \end{array}
  \right.
\end{equation}
Note also that, for each Young diagram of $N$ boxes there are $(N-1)!$
Young tableaux with the number 1 occupying the bottom left corner.
There is yet another factor of $N^2$ arising from the fact that there
are $N^2$ poles in the moduli space of flat $SU(N)$ gauge fields
corresponding to the same Young tableau. Namely, if there is a pole
corresponding to a Young tableau at $w_i=w_i^\circ~(i=1,\cdots,N)$, then
there are actually $N^2$ poles corresponding to the same tableau at
\begin{equation}
 w_i=w_i^\circ + \frac{k+l\tau}N\,.~~\Big( k,l\in\{0,1,\cdots,N-1\}\Big)
\end{equation}
The product of these factors $N^{-1}\cdot (N-1)!\cdot N^2$ precisely
cancels with the order of Weyl group in the denominator of
(\ref{Z1L-ABJM}). Thus one can show, assuming that other poles not
corresponding to any Young tableau do not contribute, that the
Jeffrey-Kirwan residue of $Z_\text{1-loop}$ (\ref{Z1L-ABJM}) reproduces
the Young diagram sum (\ref{EG1}) with $2m=\epsilon_1+\epsilon_2$ substituted.

Let us now think of more general arrangements ${\bf A}$ of $N$ particles
on a lattice, not necessarily corresponding to Young tableaux.
Consider the Jeffrey-Kirwan residue of the following meromorphic form
\begin{equation}
 \hat Z_\text{1-loop}({\bf A}) = {\rm d}^{N-1}w\cdot
\frac{\displaystyle
 \prod_{\scriptscriptstyle(\cdot\!\cdot)}(w_j-w_i)^2
 \prod_{\scriptscriptstyle\nearrow}(w_j-w_i-\epsilon_1-\epsilon_2)}
{\displaystyle
 \prod_{\scriptscriptstyle\rightarrow}(w_j-w_i-\epsilon_1)\cdot
 \prod_{\scriptscriptstyle\uparrow}(w_j-w_i-\epsilon_2)},
\label{ZhatA}
\end{equation}
which is the simplified version of $Z_\text{1-loop}$ at the pole
corresponding to ${\bf A}$. The first factor in the enumerator is the
product of double zeroes for pairs of particles occupying the same
lattice site.

We apply to this meromorphic form the general formula for Jeffrey-Kirwan
residue given at the beginning of this section. First, $\Delta$ in this
case is the set of all the links connecting neighboring particles. Then
the bases of $\Delta$ are identified with tree graphs, though now they
are allowed to have overlapping particles or links.
For arbitrary choice of the set $B$ of bases of $\Delta$, one should be
able to write $\hat Z_\text{1-loop}({\bf A})$ as a sum of derivatives of
basic fractions $\phi_{b~(b\in B)}$ and fractions with trivial
residue. The Jeffrey-Kirwan residue of the basic fraction $\phi_b$ is
nonvanishing only when the corresponding tree graph is monotonic. In
order for $\Delta$ to have at least one basis corresponding to a
monotonic tree graph, the corresponding particle arrangement ${\bf A}$
must fit within the first quadrant and the 1-st particle has to sit at
its bottom-left corner. Some simple examples of such ${\bf A}$ are those
given in (\ref{PA-YD}) and (\ref{PA-1stq}). Let us restrict our argument
to such arrangements in what follows.

We arrange the elements of $\Delta$ so that the height of the
corresponding links are non-decreasing. (Of course, this requirement
does not fix the order uniquely.) Then, with a suitable choice of the
basis set $B$, one can express the Jeffrey-Kirwan residue as a linear
sum of iterated residues $\text{Res}_b$, where the sum runs only over
those $b\in B$ corresponding to monotonic tree graphs. In all of these
iterated residues, one first uses the translation invariance of
$\hat Z_\text{1-loop}$ to fix $w_1$ corresponding to the 1-st particle at the
corner site of height $0$, then integrate the $w_i$'s corresponding to
the particles at height $1$, then those corresponding to the height $2$
and so on. This procedure can be viewed as putting one box after another
in the first quadrant. If at each step of iteration a box is put in an
incorrect place that violates the rule of making a Young diagram, the
residue of $\hat Z_\text{1-loop}({\bf A})$ vanishes due to the effect of
its enumerator. Thus the only contribution to the elliptic genus is from
the poles corresponding to Young tableaux with the number $1$
at the bottom left corner.

\section{Concluding Remarks}\label{sec:conclusion}

Our analysis of the ABJM model with boundary was able to reproduce the
known observables for multiple self-dual strings only partially, and
there remains a significant mismatch. We believe that it should be
resolved by a deeper understanding of the ABJM model on the boundary
and the IIA brane model.

In making more precise comparisons between the two descriptions through
elliptic genera or other $R$-independent physical observables, one also
needs to be careful about what those observables really are. A good
example where such a subtlety arises is a microscopic derivation of the
quarter-BPS-dyon counting formula in ${\cal N}=4$ string theory using
the 4d-5d lift \cite{Gaiotto:2005gf}. We start with the D1-D5 system in
IIB on $K3\times S^1\times$Taub-NUT. When the radius $R$ of the asymptotic
circle of the Taub-NUT space is large, the above system describes the
Strominger-Vafa black hole with angular momentum. When $R$ is small, the
system can be related to a quarter BPS dyon using the string
duality. Thus one can expect that the partition function of these dyons,
independent of $R$, equals to that of the rotating 
Strominger-Vafa black holes. The former is given by the so-called Igusa
form and the latter by the elliptic genus of symmetric products of
$K3$. Indeed they are very similar but not exactly the same. To have the
exact match, it turns out that one also needs to consider other
contributions such as the center of mass motion of the D1-D5 system in
the Taub-NUT space and bound states of momentum along $S^1$ with the KK
monopole \cite{David:2006yn}.

The above example suggests that, in order to have an exact match between
the elliptic genera of the ABJM slab and the IIA brane model, we may also
need to take care of those additional contribution such as the c.o.m
motion of the M2-M5 system in the Taub-NUT space. We leave this for the
future. 

In the present work, we studied the simplest boundary condition relevant
to  the worldsheet theory of the self-dual string, and did not consider
possible boundary degrees of freedom such as those proposed in
\cite{Chu:2009ms}. It would be very
interesting to fully explore the $1/2$ BPS supersymmetric boundary
conditions of the ABJM model along the line of \cite{Gaiotto:2008sa}. In
addition it is important to generalize our analysis to a system of
M2-branes intersecting with M5-branes, i.e., domain wall to construct
the quiver models discussed in section 4.4.

Another direction to study further is a generalization to ${\cal N}=5$
Chern-Simons-Matter theories \cite{Hosomichi:2008jb,Aharony:2008gk} on
the boundary and interval, and other boundary conditions such as from
the 9-branes \cite{Haghighat:2014pva}. 

\section*{Acknowledgments}

We would like to thank Kentaro Hori, Seok Kim, David Kutasov, Emil
Martinec, Hirosi Ooguri and Soo-Jong Rey for discussions. We thank the
organizers of the Simons Summer Workshop 2013 and the KITP program ``New
Methods in Nonperturbative Quantum Field Theory'' where part of this
work has been carried out. KH also thanks the organizers of the Kavli
IPMU-FMSP Workshop ``Supersymmetry in Physics and Mathematics'' for
hospitality. The work of KH is supported in part by MEXT/JSPS
Grant-in-Aid for Scientific Research No. 26400247. The work of SL is
supported in part by the Ernest Rutherford fellowship of the Science \&
Technology Facilities Council ST/J003549/1.

\appendix

\section{$\bf U(1)\times U(1)$ ABJM model on a slab}\label{sec:appendix}

In order to study SUSY-protected sector of the worldsheet theory of
self-dual strings, it may be essential to formulate the ABJM model on a
slab in an off-shell supersymmetric manner including Kaluza-Klein degree
of freedom. We need to reorganize the fields into multiplets of 2D SUSY,
and regard each field as carrying an additional continuously-varying
label $y$ in a way similar to \cite{Gaiotto:2008sa,Gaiotto:2008sd}. Here
we illustrate this procedure for the simplest example of Euclidean
$U(1)\times U(1)$ ABJM theory. The Lagrangian on flat $\mathbb R^3$ is
\begin{eqnarray}
 {\cal L} &=& \frac{ik}{4\pi}\varepsilon^{mnp}
 (A_m\partial_nA_p-\tilde A_m\partial_n\tilde A_p)
 +D_m\bar Z^aD^mZ_a-\bar\Psi_a\gamma^mD_m\Psi^a
 \nonumber \\ &=&
 \frac{ik}{2\pi}\varepsilon^{mnp}B_m\partial_nC_p
 +D_m\bar Z^aD^mZ_a-\bar\Psi_a\gamma^mD_m\Psi^a,
\label{Lu1}
\end{eqnarray}
where $B_m\equiv\frac12(A_m+\tilde A_m),\, C_m\equiv A_m-\tilde A_m$ and
the covariant derivative of matters is $D_mZ_a=(\partial_m-iC_m)Z_a$ etc.
We would like to put it on $S^2\times(\text{interval})$ preserving 2D
${\cal N}=(2,2)$ off-shell supersymmetry associated with the Killing
spinor on the round $S^2$ of radius $\ell$,
\begin{equation}
 D_\mu\epsilon=\frac{i}{2\ell}\gamma_\mu\epsilon.
\end{equation}
We use the same convention for the spinor calculus as was summarized at
the beginning of section \ref{sec:ABJM}, except $\gamma^m~(m=1,2,3)$
here are chosen to be Pauli's matrices. We follow the general
construction of ${\cal N}=(2,2)$ theories on $S^2$ given in
\cite{Benini:2012ui,Doroud:2012xw}.

The boundary conditions on various fields are
\begin{eqnarray}
 \text{Neumann}&:&
 B_\mu,~~ C_3,~~
 Z_I,~\Psi^I_-,~
 \Psi^A_+,
 \nonumber \\
 \text{Dirichlet}&:&
 C_\mu,~~ B_3,~~
 Z_A,~\Psi^A_-,~
 \Psi^I_+.\quad(\mu=1,2;\,I=1,2;\,A=3,4)
\end{eqnarray}
Note that the fields in the same multiplet must obey the same boundary
condition.

Let us first focus on the terms in (\ref{Lu1}) involving the covariant
$x_\mu$-derivative of matters. $C_\mu$ is in an abelian vector multiplet,
\begin{eqnarray}
 2\delta C_\mu &=& \xi\gamma_\mu\bar\lambda+\bar\xi\gamma_\mu\lambda,
 \nonumber \\
 2\delta\sigma_3 &=& \xi\gamma_3\bar\lambda+\bar\xi\gamma_3\lambda,
 \nonumber \\
 2\delta\sigma_4 &=& i\xi\bar\lambda-i\bar\xi\lambda,
 \nonumber \\
 \delta\lambda &=& i\gamma^3\xi F_{12}+\xi D
  +\gamma^{\mu3}D_\mu(\xi\sigma_3)
 -i\gamma^\mu D_\mu(\xi\sigma_4),
 \nonumber \\
 \delta\bar\lambda &=& i\gamma^3\bar\xi F_{12}-\bar\xi D
  +\gamma^{\mu3}D_\mu(\bar\xi\sigma_3)
 +i\gamma^\mu D_\mu(\bar\xi\sigma_4),
 \nonumber \\
 2\delta D &=& D_\mu(\xi\gamma^\mu\bar\lambda)-D_\mu(\bar\xi\gamma^\mu\lambda),
\end{eqnarray}
Here $F_{12}$ is the field strength of $C_\mu$.
The matter fields are organized into four chiral multiplets with the
lowest components $\phi_i~(i=1,\cdots,4)$, which couple to $C_\mu$
according to the charges $e_i=(+1,-1,+1,-1)$. Their transformation rule reads
\begin{eqnarray}
 \delta\phi_i &=& \xi\psi_i,
 \nonumber \\
 \delta\bar\phi_i &=& \bar\xi\bar\psi_i,
 \nonumber \\
 \delta\psi_i &=& -\gamma^\mu\bar\xi D_\mu\phi_i
 +ie_i\gamma^3\bar\xi\sigma_3\phi_i-e_i\bar\xi\sigma_4\phi_i
 +\xi F_i-\frac{q_i}2\gamma^\mu D_\mu\bar\xi\phi_i,
 \nonumber \\
 \delta\bar\psi_i &=& -\gamma^\mu\xi D_\mu\bar\phi_i
 -ie_i\gamma^3\xi\sigma_3\bar\phi_i-e_i\xi\sigma_4\bar\phi_i
 -\bar\xi\bar F_i-\frac{q_i}2\gamma^\mu D_\mu\xi\bar\phi_i,
 \nonumber \\
 \delta F_i &=& -\bar\xi\gamma^\mu D_\mu\psi_i
 +ie_i\bar\xi\gamma^3\sigma_3\psi_i+e_i\bar\xi\sigma_4\psi_i
 +ie_i\bar\xi\bar\lambda\phi_i-\frac{q_i}2D_\mu\bar\xi\gamma^\mu\psi_i,
 \nonumber \\
 \delta\bar F_i &=& +\xi\gamma^\mu D_\mu\bar\psi_i
 +ie_i\xi\gamma^3\sigma_3\bar\psi_i-e_i\xi\sigma_4\bar\psi_i
 +ie_i\xi\lambda\bar\phi_i+\frac{q_i}2D_\mu\xi\gamma^\mu\bar\psi_i.
\end{eqnarray}
Here $q_i$ are the vector R-charges. The component fields in these
multiplets are identified with $Z_a,\Psi^a$ as follows,
\begin{eqnarray}
 \phi_i &=& (Z_1,\bar Z^2,Z_3,\bar Z^4),
 \nonumber \\
 \psi_{i+} &=& (\Psi^3_+,-\bar\Psi_{4+},-\Psi^1_+,\bar\Psi_{2+}),
 \nonumber \\
 \psi_{i-} &=& (\Psi^2_-,-\bar\Psi_{1-},\Psi^4_-,-\bar\Psi_{3-}).
\end{eqnarray}
The above transformation rules are the same as the one for 2D
${\cal N}=(2,2)$ SUSY theories on $S^2$, except that all the fields now
depend also on $x^3$. The standard 2D SUSY construction gives a part of
the matter kinetic term,
\begin{eqnarray}
 {\cal L}_\text{kin} &=&
 D_\mu\bar\phi_iD^\mu\phi_i+\bar\phi_ie^2_i(\sigma_3^2+\sigma_4^2)\phi_i
 -\bar\psi_i(\gamma^\mu D_\mu-ie_i\gamma^3\sigma_3-e_i\sigma_4)\psi_i
 +\bar F_iF_i
 \nonumber \\ &&
 +e_i\Big(i\bar\phi_i D\phi_i
 -i\bar\phi_i\lambda\psi_i+i\bar\psi_i\bar\lambda\phi_i
 +\frac{iq_i}\ell\bar\phi_i\sigma_4\phi_i\Big)
 +\frac{iq_i}{2\ell}\bar\psi_i\psi_i
 +\frac{q_i(2-q_i)}{4\ell^2}\bar\phi_i\phi_i,
\end{eqnarray}
where the summation over $i$ is understood.

The other part of the kinetic term, which involves covariant
$x_3$-derivative of matter fields, should arise from F-term.
We first introduce the chiral multiplet containing $C_3$,
\begin{eqnarray}
&& \delta(C_3+iC_4) ~=~ \xi\zeta,
 \nonumber \\
&& \delta(C_3-iC_4) ~=~ \bar\xi\bar\zeta,
 \nonumber \\
&& \delta\zeta ~=~ \gamma^\mu\bar\xi(\partial_3C_\mu-\partial_\mu(C_3+iC_4))
 +\gamma^3\bar\xi\partial_3\sigma_3+i\bar\xi\partial_3\sigma_4
 +\xi H,
 \nonumber \\
&& \delta\bar\zeta ~=~ \gamma^\mu\xi(\partial_3C_\mu-\partial_\mu(C_3-iC_4))
 +\gamma^3\xi\partial_3\sigma_3-i\xi\partial_3\sigma_4
 +\bar\xi\bar H,
 \nonumber \\
&& \delta H ~=~
 -\bar\xi\gamma^\mu D_\mu\zeta+\bar\xi\partial_3\bar\lambda,
 \nonumber \\
&& \delta\bar H ~=~
 +\xi\gamma^\mu D_\mu\bar\zeta-\xi\partial_3\lambda.
\end{eqnarray}
The above transformation rule is that of $\partial_3\cdot\log$ of an
ordinary charged chiral multiplet, which ensures that $C_3$ transforms as
the third component of the gauge field $C$ under gauge transformations.
From the gauge-invariant superpotential,
\begin{equation}
 W=\phi_2(\partial_3-iC_3+C_4)\phi_3-\phi_4(\partial_3-iC_3+C_4)\phi_1,
\end{equation}
we obtain the following F-term invariant
\begin{eqnarray}
\lefteqn{
 {\cal L}_\text{F-term} ~=~
   iH(\phi_1\phi_4-\phi_2\phi_3)
  +i\bar H(\bar\phi_1\bar\phi_4-\bar\phi_2\bar\phi_3)
}
 \nonumber \\ &&
 -F_4(D_3\phi_1+C_4\phi_1)
 -F_3(D_3\phi_2-C_4\phi_2)
 +F_2(D_3\phi_3+C_4\phi_3)
 +F_1(D_3\phi_4-C_4\phi_4)
 \nonumber \\ &&
 +\bar F_4(D_3\bar\phi_1+C_4\bar\phi_1)
 +\bar F_3(D_3\bar\phi_2-C_4\bar\phi_2)
 -\bar F_2(D_3\bar\phi_3+C_4\bar\phi_3)
 -\bar F_1(D_3\bar\phi_4-C_4\bar\phi_4)
 \nonumber \\ &&
 +\psi_4(D_3+C_4)\psi_1
 -\psi_2(D_3+C_4)\psi_3
 +\bar\psi_4(D_3+C_4)\bar\psi_1
 -\bar\psi_2(D_3+C_4)\bar\psi_3
 \nonumber \\ &&
 -i\zeta(\phi_4\psi_1+\psi_4\phi_1-\phi_2\psi_3-\psi_2\phi_3)
 +i\bar\zeta(\bar\phi_4\bar\psi_1+\bar\psi_4\bar\phi_1
    -\bar\phi_2\bar\psi_3-\bar\psi_2\bar\phi_3).
\end{eqnarray}
Note that, due to the superpotential, the R-charges of matter chiral
multiplet must satisfy $q_1+q_4=q_2+q_3=2$. In addition, if we are
putting nonzero classical values to $\phi_3=Z_3$ and $\phi_4=\bar Z^4$,
the supersymmetry will be broken unless $q_3=q_4=0$.

Let us introduce another gauge field $B_m$ and construct Chern-Simons
Lagrangian. The first two components $B_\mu$ belong to a chiral
multiplet of vector R-charge $q=2$ or ``twisted vector multiplet,
\begin{eqnarray}
&& 2\delta B_\mu ~=~ \xi\gamma_{\mu3}\bar\chi+\bar\xi\gamma_{\mu3}\chi,
 \nonumber \\
&& \delta(\rho_3+i\rho_4) ~=~ \xi\chi,
 \nonumber \\
&& \delta(\rho_3-i\rho_4) ~=~ \bar\xi\bar\chi,
 \nonumber \\
&& \delta\chi ~=~ -i\gamma^\mu D_\mu\big(\bar\xi(\rho_3+i\rho_4)\big)
 +\xi(E+iG_{12}),
 \nonumber \\
&& \delta\bar\chi ~=~ -i\gamma^\mu D_\mu\big(\xi(\rho_3-i\rho_4)\big)
 -\bar\xi(E-iG_{12}),
 \nonumber \\
&& \delta(E+iG_{12}) ~=~ -D_\mu(\bar\xi\gamma^\mu\chi),
 \nonumber \\
&& \delta(E-iG_{12}) ~=~ +D_\mu(\xi\gamma^\mu\bar\chi),
\end{eqnarray}
where $G_{12}$ is the field strength of $B_\mu$.
$B_3$ belongs to a twisted chiral multiplet of axial R-charge 0 which
couples to the above twisted vector multiplet in the same way the chiral
multiplet $C_3$ couples to the vector multiplet $C_\mu$.
\begin{eqnarray}
 2\delta B_3 &=& \xi\gamma_3\bar\eta+\bar\xi\gamma_3\eta,
 \nonumber \\
 2\delta B_4 &=& i\xi\bar\eta-i\bar\xi\eta,
 \nonumber \\
 \delta\eta &=& i\gamma^3\xi K_4+\xi K_3
 +\gamma^{\mu3}\xi D_\mu B_3
 -i\gamma^\mu\xi D_\mu B_4
 +\bar\xi\partial_3(\rho_3+i\rho_4)-\gamma^{\mu3}\xi\partial_3B_\mu,
 \nonumber \\
 \delta\bar\eta &=& i\gamma^3\bar\xi K_4-\bar\xi K_3
 +\gamma^{\mu3}\bar\xi D_\mu B_3
 +i\gamma^\mu\bar\xi D_\mu B_4
 +\xi\partial_3(\rho_3-i\rho_4)-\gamma^{\mu3}\bar\xi\partial_3B_\mu,
 \nonumber \\
 2\delta K_3 &=& \xi\gamma^\mu D_\mu\bar\eta-\bar\xi\gamma^\mu D_\mu\eta
 -\bar\xi\partial_3\chi+\xi\partial_3\bar\chi,
 \nonumber \\
 2\delta K_4 &=& i\xi\gamma^{\mu3}D_\mu\bar\eta+i\bar\xi\gamma^{\mu3} D_\mu\eta.
 +\bar\xi\gamma_3\partial_3\chi+\xi\gamma_3\partial_3\bar\chi.
\end{eqnarray}
If we temporarily forget about the gauge non-invariance of $B_3,C_3$ and
apply the standard construction of 2D F-term and twisted F-term, we
obtain the following Chern-Simons like Lagrangian,
\begin{eqnarray}
 {\cal L}_\text{cs} &=&
 \frac{ik}{2\pi}(F_{12}B_3+DB_4+ K_4\sigma_3+K_3\sigma_4
                +G_{12}C_3+EC_4+ H_4\rho_3 + H_3\rho_4)
 \nonumber \\ &&
 +\frac k{4\pi}(\lambda\bar\eta+\bar\lambda\eta
                -\chi\zeta-\bar\chi\bar\zeta),
\end{eqnarray}
where we used $H=H_3+iH_4, \bar H=H_3-iH_4$. This ${\cal L}_\text{cs}$
is actually not supersymmetric, but it is cured by adding
\begin{equation}
 \delta{\cal L}_\text{cs} =
 \frac{ik}{2\pi}(B_2\partial_3C_1-B_1\partial_3C_2),
\end{equation}
which is also the right term to complete the Chern-Simons Lagrangian.

\end{document}